\documentclass[prd,singlecolumn]{revtex4}
\pdfoutput=1
\usepackage{amssymb,latexsym}
\usepackage{amsmath,amsbsy,bbm}
\usepackage{ifpdf}
\usepackage{epsfig,bm}
\usepackage{graphicx,comment}
\usepackage{color}
\usepackage{soul}
\usepackage{mathtools}
\usepackage{comment}
\usepackage[normalem]{ulem}
\begin{document}

\title{Phase transitions of the antiferromagnetic $J_1$-$J_2$-$J_3$ Ising model on the cubic lattice}
\author{Kai-Wei Huang}
\affiliation{Department of Physics, National Taiwan Normal University,
	88, Sec.4, Ting-Chou Rd., Taipei 116, Taiwan}
\author{Chien-Ting Chen}
\affiliation{Department of Physics, National Taiwan Normal University,
	88, Sec.4, Ting-Chou Rd., Taipei 116, Taiwan}	
\author{Shang-Wei Li}
\affiliation{Department of Physics, National Taiwan Normal University,
88, Sec.4, Ting-Chou Rd., Taipei 116, Taiwan}
\author{Fu-Jiun Jiang}
\email[]{fjjiang@ntnu.edu.tw}
\affiliation{Department of Physics, National Taiwan Normal University,
88, Sec.4, Ting-Chou Rd., Taipei 116, Taiwan}

\begin{abstract}
Motivated by the conflicting findings in prior studies regarding the nature of the phase
transition for the three-dimensional antiferromagnetic $J_1$-$J_2$-$J_3$ Ising model on the cubic lattice with $J_1 = J_2 = J_3 = 1.0$, in this study we conduct a detailed investigation of this model using both Monte Carlo simulations and neural networks (NN). Here $J_1$ and $J_2$
are the intralayer nearest-neighbor and next-nearest-neighbor couplings, respectively, and $J_3$
is the nearest-neighbor interlayer interaction. By investigating the data obtained from much larger lattices than those used in previous 
studies, we find that the phase transition for $J_1 = J_2 = J_3 = 1.0$ is second order 
without any ambiguity. Hence we bring a resolution to this debate. Apart from $J_1 = J_2 = J_3 = 1.0$, we study the critical phenomenon associated with $J_2/J_1 = 4/7$ and $J_3 = J_1 = 1.0$ as well. In addition, by employing a dedicated training and testing strategy, our NN computation not only is much more efficient than those utilizing the standard procedures commonly used in the literature, but also requires only a per-mille fraction of the storage capacity needed by a conventional NN calculation when a large linear system size $L$ is considered. The method employed here provides an extremely resource-efficient route for applying NN techniques to study the critical behaviors of higher-dimensional systems.    
  
\end{abstract}

\maketitle

\section{Introduction}\vskip-0.3cm

The study of classical spin systems with competing interactions is one of the central research directions in statistical mechanics \cite{Die13}. Two-dimensional (2D) frustrated Ising models with ferromagnetic nearest-neighbor coupling $J_1$ and antiferromagnetic next-nearest-neighbor coupling $J_2$ on both the square and the honeycomb lattices are two such examples \cite{Oit87,Lop93,Lop931,Kal08,Jin12,Kal12,Jin13,Kel19,Hu21,Li21,Yos23,Li24,Bob16,Zuk19,Zuk21,Ace21,Sch21,Ges24,Roo24,Azh25,Bat26,Dia26,Ges26,Li26}. Many efforts have been devoted to uncovering the critical phenomena of these models. As a result, the phase diagrams of the frustrated Ising model on both the square and the honeycomb lattices are well determined, although some details regarding the nature and the critical exponents of the critical phenomena of the aforementioned physical systems are still under debate \cite{Kal08,Jin12,Kal12,Jin13,Li21,Yos23,Li24,Azh25,Li26}. For instance, the tricritical point in the parameter space of the frustrated Ising model on the honeycomb lattice, where the phase transitions change from second order to first order, remains to be determined with high precision \cite{Azh25,Li26}.        

Like the frustrated $J_1$-$J_2$ Ising model on the square lattice, the 3D analog, namely
the antiferromagnetic $J_1$-$J_2$-$J_3$ Ising model on the cubic lattice, has been investigated as well.
Here $J_1 > 0$ and $J_2 > 0$ are the intralayer nearest-neighbor and next-nearest-neighbor couplings, respectively. In addition, $J_3 > 0$ is the interlayer coupling between any two consecutive layers.
It is anticipated that as the ratio $J_3/J_1$ decreases, the phase diagram of the 2D $J_1$-$J_2$ Ising model
can be recovered. Therefore, the 3D model may serve as another route to explore the phase diagram of
the 2D model. Apart from this, studying the 3D model on the cubic lattice itself is an intriguing research topic as well.

The case of $J_1 = J_2 = J_3 = 1.0$ was studied in Ref.~\cite{Ram15} using the Metropolis algorithm with the technique of parallel tempering \cite{Met53,Huk96}. In that study it was observed that   
a two-peak distribution in the energy density $E$ appeared on small lattices, but was gone
for the linear system sizes $L=64$ and $90$. As a consequence, the targeted phase transition
is pseudo-first order. Later, a study of the same 3D model with $J_1=J_2=J_3 = 1.0$ using the Wang-Landau algorithm \cite{Wan01} found that a two-peak
structure showed up in the histogram of $E$ for $L=90$, and claimed that the considered phase transition is first order \cite{Sor18}. Finally, an investigation conducted in the year 2022 using the Metropolis algorithm concluded that the aforementioned phase transition is first order based on the histograms of $E$ with $L \le 128 $ \cite{Sor22}. 
  
Due to the controversial results pointed out above regarding the nature of the phase transition of the 3D antiferromagnetic model with isotropic couplings, in this study we investigate the targeted critical behavior by utilizing both the methods of Monte Carlo simulations (MC) and neural networks (NN).
In particular, the Monte Carlo computations are performed using the Metropolis algorithm with parallel tempering, and an unconventional approach is employed for the NN calculations.
Apart from the isotropic case, the case with $J_2/J_1 = 4/7$ and $J_3=J_1 = 1.0$ is considered here as well.

Remarkably, by simulating much larger lattices than those considered in previous studies,
we obtain convincing evidence that the phase transition for $J_1 = J_2 = J_3 = 1.0$
is second order based on the histograms of $E$, and therefore resolve the long-standing controversial
issue. We also find that for the case of $J_2 = 4/7$ and $J_1 = J_3 = 1.0$, which is known to be a strong first-order phase transition, the phenomenon of energy tunneling may not be exponentially suppressed as $\exp(-c\beta L^{2})$ ($\beta$ is the inverse temperature and $c$ is determined by the interfacial tension) if one of the ground-state stripe configurations is used as the initial configuration to start the MC simulations. This is a surprising observation and is worth exploring further. 

Moreover, with an unusual NN training approach, only subsets, which are a per-mille fraction of the whole configurations,
are needed to reach correct determinations of the nature of the phase transitions for $J_2 = 1.0$ and $J_2 = 4/7$ with $J_1 = J_3 = 1.0$. The idea used here not only saves a huge amount of storage space but also boosts our NN computations dramatically when compared to the standard procedures commonly employed in the literature. The NN approach utilized here is perfectly suited for conducting NN calculations for
higher-dimensional systems and outperforms the conventional NN methods.

In conclusion, our study brings a solution to the controversial issue regarding the nature of the phase transition of the considered 3D antiferromagnetic Ising model with $J_1 = J_2 = J_3 = 1.0$,
and the strategy used in the NN computation presented here provides a much more efficient approach
than the standard NN procedures when higher-dimensional systems are investigated using the NN methods.

The rest of the paper is summarized as follows. After the brief introduction, the model and the calculated quantities are described in Sect. II.
We then detail our NN ideas and present both the MC and NN numerical data, which support the claims given in the Introduction, in
Sect. III. Finally, in Sect. IV, we summarize our findings of this investigation.

\section{The considered model}

The Hamiltonian $H$ of the 3D frustrated $J_1$-$J_2$-$J_3$ Ising model on the cubic
lattice studied here is expressed as \cite{Jin12}
\begin{equation}
 H = J_1 \sum_{\left< ij\right>} \sigma_i\sigma_j + J_2 \sum_{\left<\left< lm\right>\right>}\sigma_l\sigma_m + J_3 \sum_{\left< st\right>} \sigma_i\sigma_j ,
\label{eqn}
\end{equation}
where $\sigma_i = \pm 1$ is the Ising spin at site $i$. Moreover, the first
term is summed over nearest-neighbor intralayer sites $i$ and $j$,
the second term is summed over next-nearest-neighbor intralayer sites $l$ and $m$, and
the third term is summed over nearest-neighbor interlayer sites $s$ and $t$. 
In our calculations, we have set $J_1 = J_3 = 1.0$ and have considered the phase
transitions for two fixed values of $J_2 > 0$, namely $J_2 = 1.0$ and $J_2 = 4/7$.

The quantities determined in this study are the energy density $E$ (energy per site) and the specific heat $C_v$, which is given by

\begin{equation}
	C_v = \frac{N}{T^2}\left( \langle E^2\rangle -\langle E\rangle^2 \right),
\end{equation}
where $N = L^3$ is the number of spins for a system with linear system size $L$.

\section{Numerical Results}

To investigate the proposed phase transitions of the 3D antiferromagnetic $J_1$-$J_2$-$J_3$ 
Ising model on the cubic lattice, we have carried out large-scale Monte
Carlo calculations (MC) using the single-spin-flip Metropolis algorithm supplemented by the parallel tempering
technique. In addition, a simple multi-layer perceptron (MLP) having one input layer, one hidden layer,
and one output layer is employed. The architecture of the MLP, the training strategy, as well as the testing procedure considered in this study will be detailed later
in the relevant parts associated with the NN computations.

\subsection{The Monte Carlo Results}

In this study, $J_3$ and $J_1$ are set to be $1.0$. In addition, we have considered two values of $J_2$, namely, $J_2=1.0$ and $J_2 = 4/7$. While the phase transition of $J_2=4/7$ is known to be first order, the nature of
the critical behavior related to $J_2 = 1$ is still controversial.

\begin{figure}	
	\begin{center}
		\hbox{
			\includegraphics[width=0.33\textwidth]{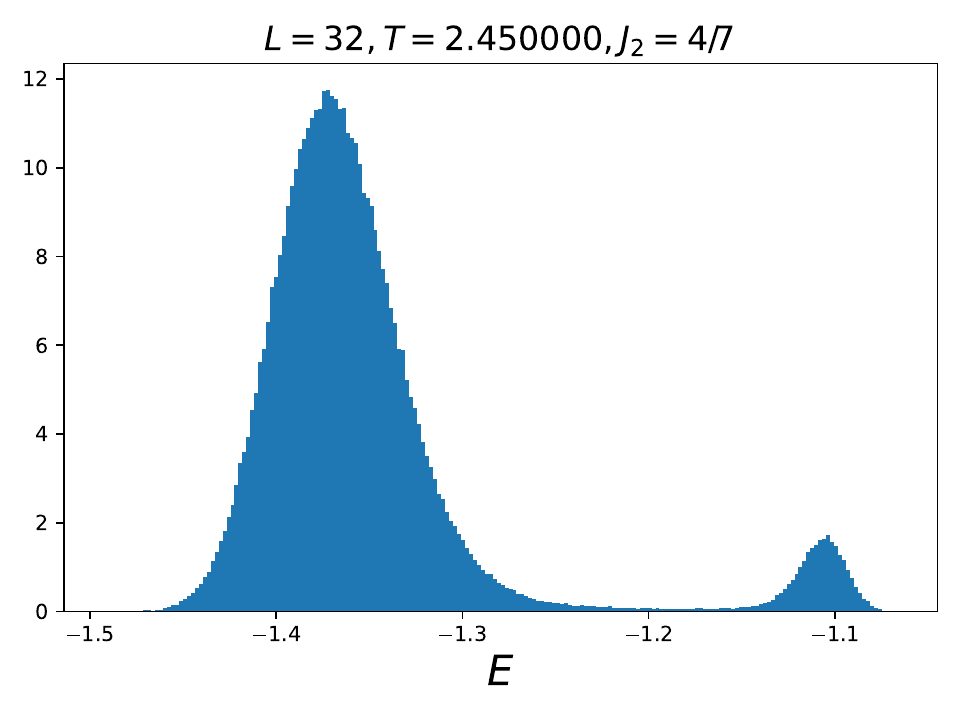}
			\includegraphics[width=0.33\textwidth]{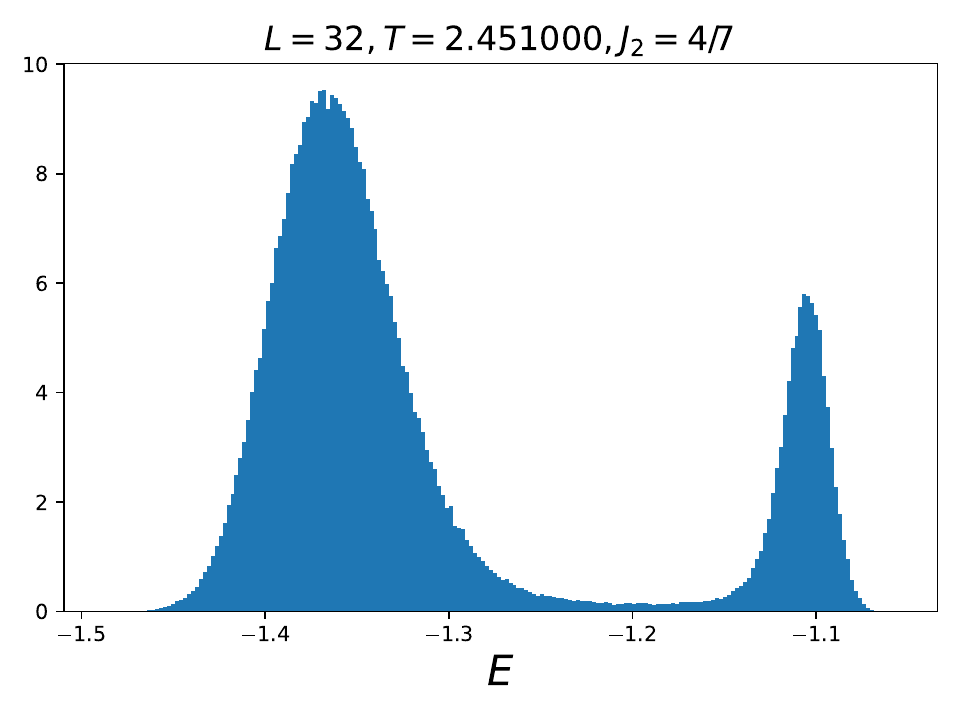}
			\includegraphics[width=0.33\textwidth]{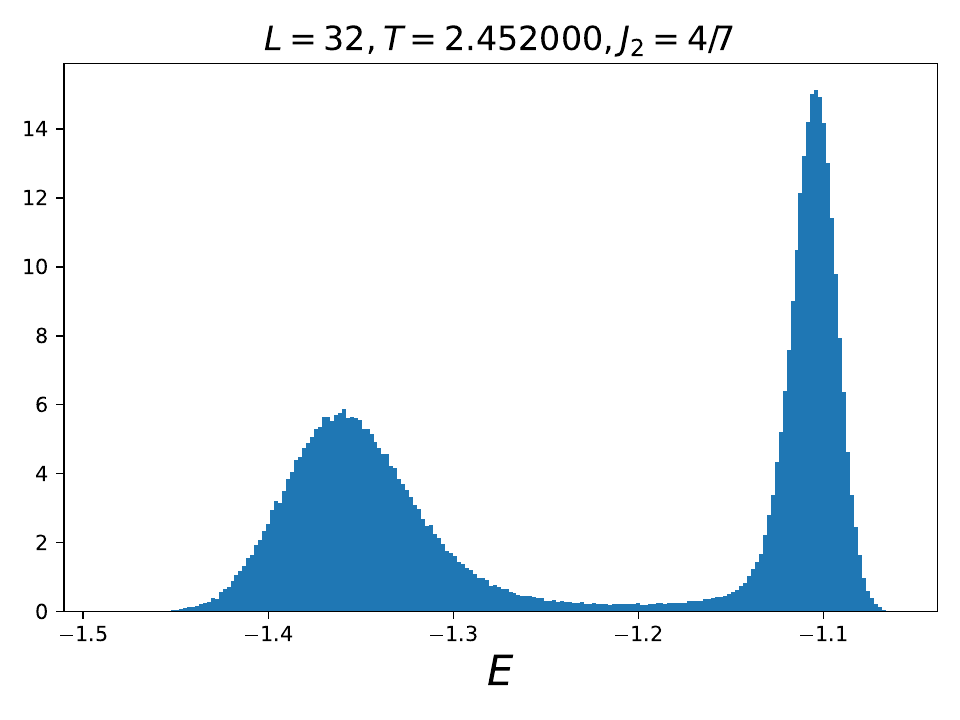}
		}
	\end{center}
	\caption{Histograms of $E$ for $J_2=4/7$ and $L=32$.
	}\label{f4/7L32}
\end{figure}

\begin{figure}

	\begin{center}
		\hbox{
			\includegraphics[width=0.33\textwidth]{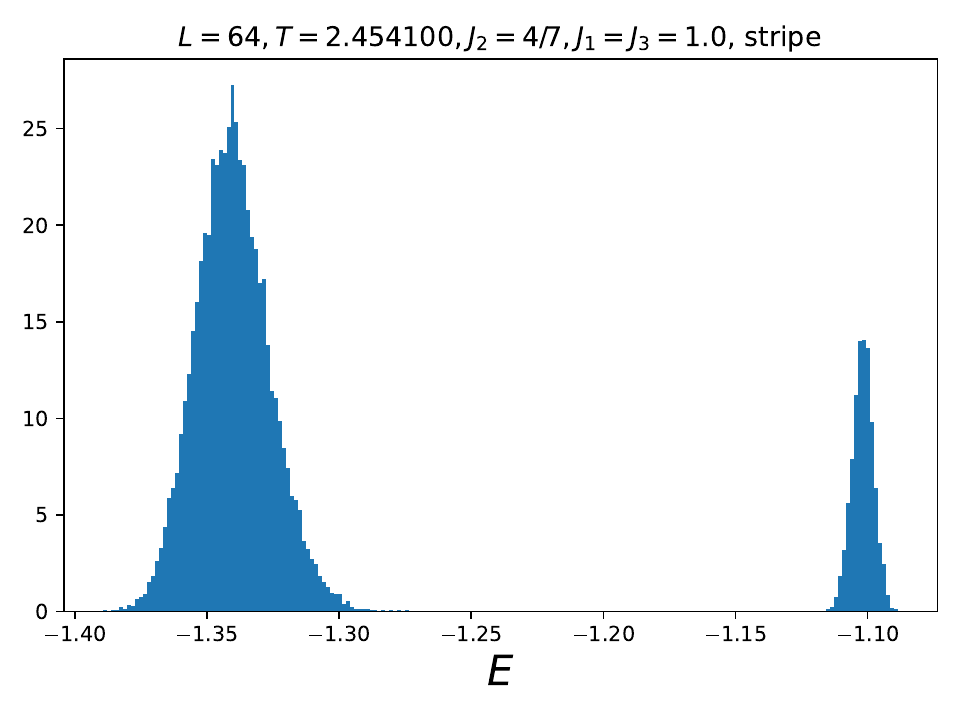}
			\includegraphics[width=0.33\textwidth]{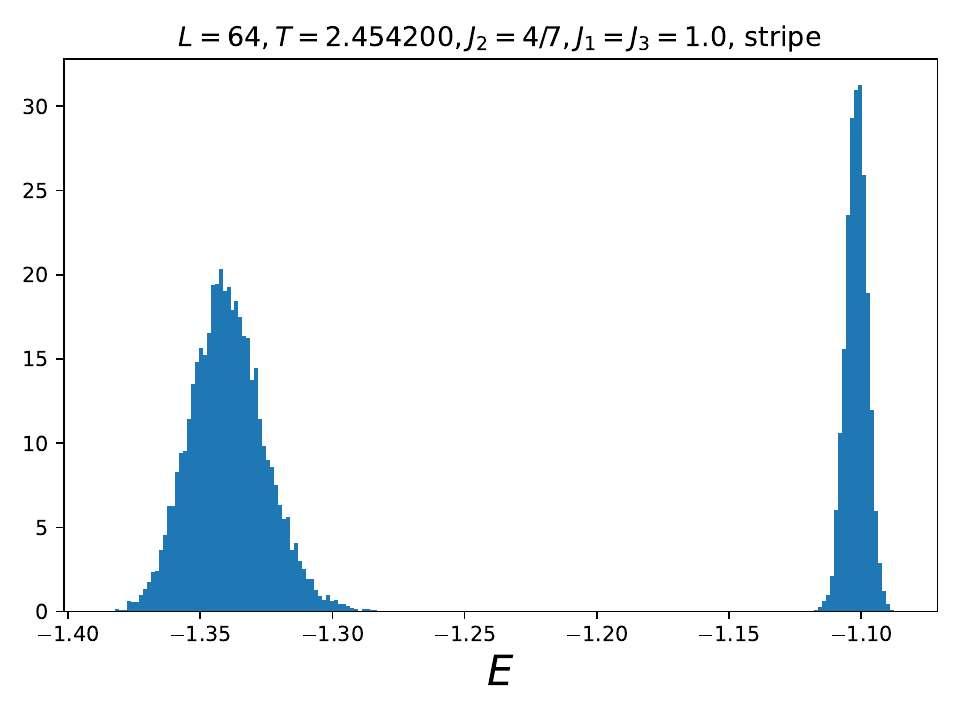}
			\includegraphics[width=0.33\textwidth]{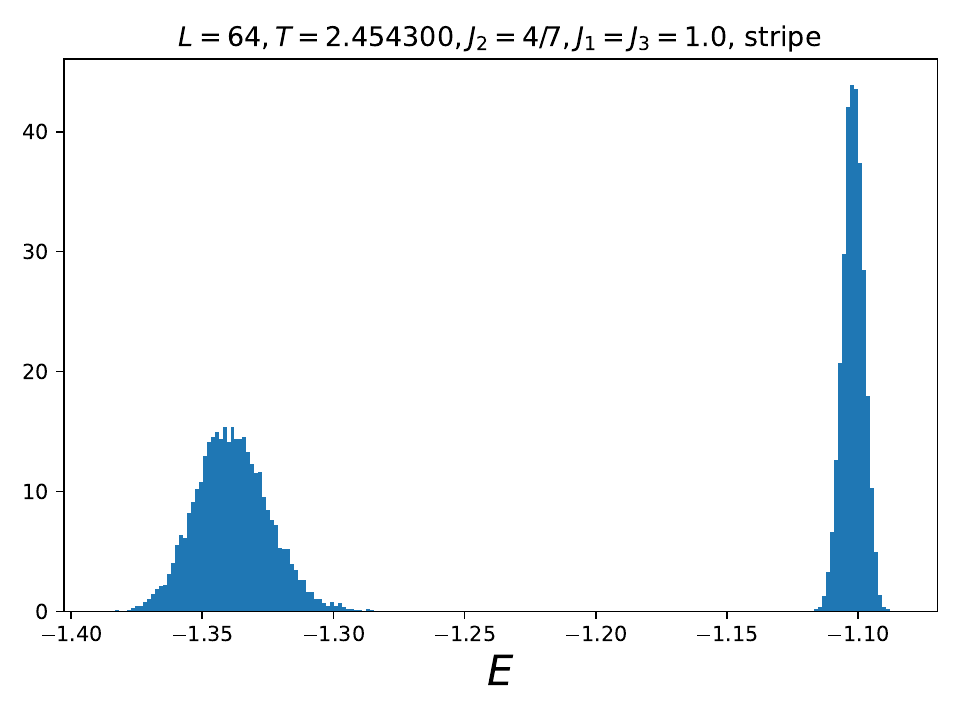}
		}
	\end{center}
	\caption{Histograms of $E$ for $J_2=4/7$ and $L=64$. The initial configuration to start the MC simulations
		is one of the theoretical ground-state stripe configurations.
	}	\label{4/7L64}
\end{figure}

\begin{figure}
	
	\begin{center}
		\hbox{
			\includegraphics[width=0.3\textwidth]{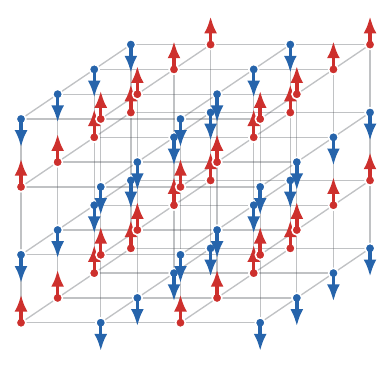}
			\includegraphics[width=0.3\textwidth]{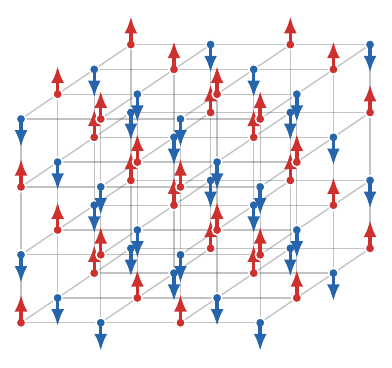}
			\includegraphics[width=0.3\textwidth]{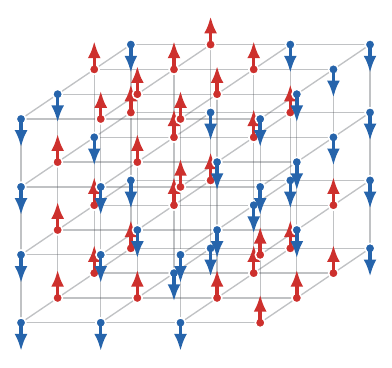}
		}
	\end{center}
	\caption{Three initial configurations to start the MC simulations. Up and down arrows stand
		for the spins having values of 1 and -1, respectively.
	}	\label{initial_conf}
\end{figure}

Fig.~\ref{f4/7L32} shows the histograms of $E$ for $J_2 = 4/7$ and $L=32$ for three different values of temperature $T$. Clearly, a two-peak structure appears in all three panels of Fig.~\ref{f4/7L32}. In addition, the histograms of $E$ for the same value of $J_2$ but with $L=64$ are given in Fig.~\ref{4/7L64}. As can be observed from Figs.~\ref{f4/7L32} and \ref{4/7L64}, the two-peak structure, which is a characteristic of a first-order
phase transition, becomes more pronounced as $L$ increases. The outcomes demonstrated in Fig.~\ref{4/7L64} are obtained from data using one of the theoretical ground-state stripe configurations as the initial configuration for the associated MC simulations (see the left panel of Fig.~\ref{initial_conf}). 

\begin{figure}
	
	\begin{center}
		\hbox{
			\includegraphics[width=0.33\textwidth]{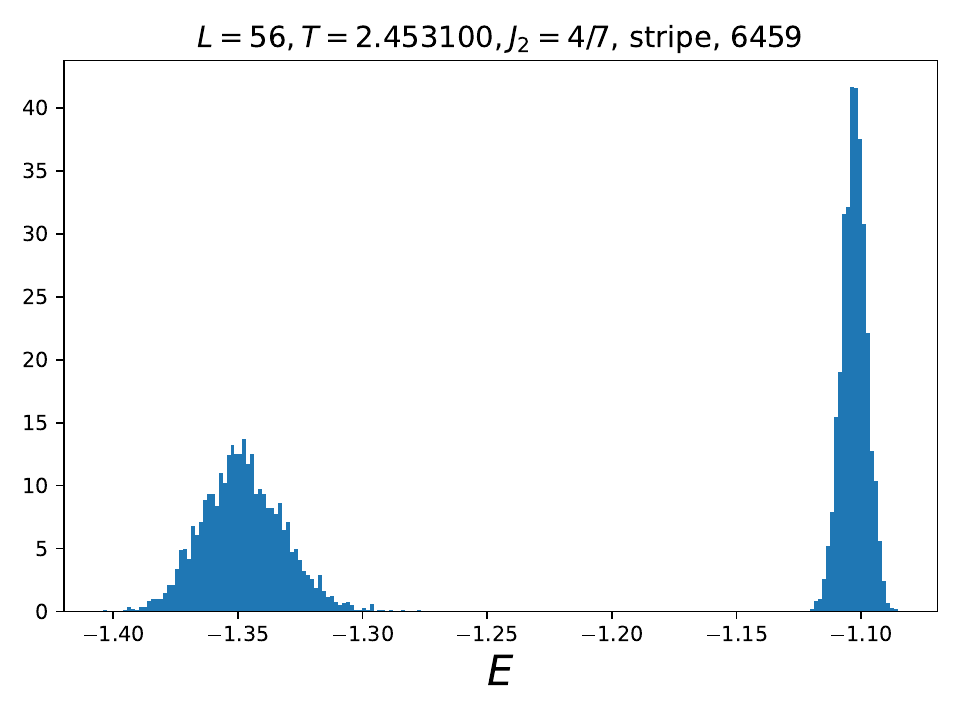}
			\includegraphics[width=0.33\textwidth]{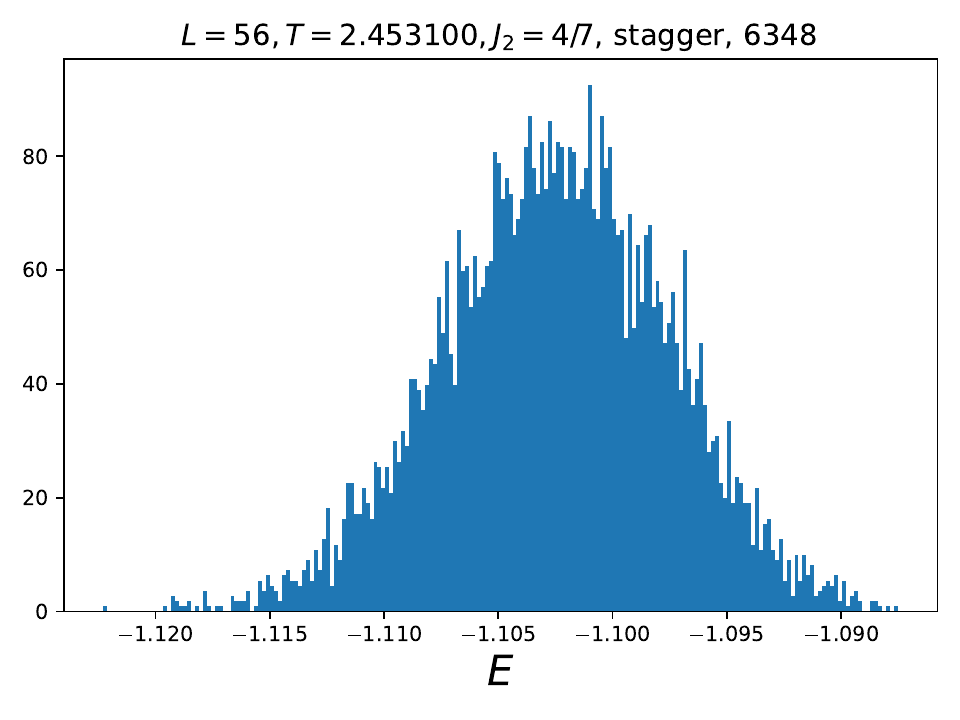}
			\includegraphics[width=0.33\textwidth]{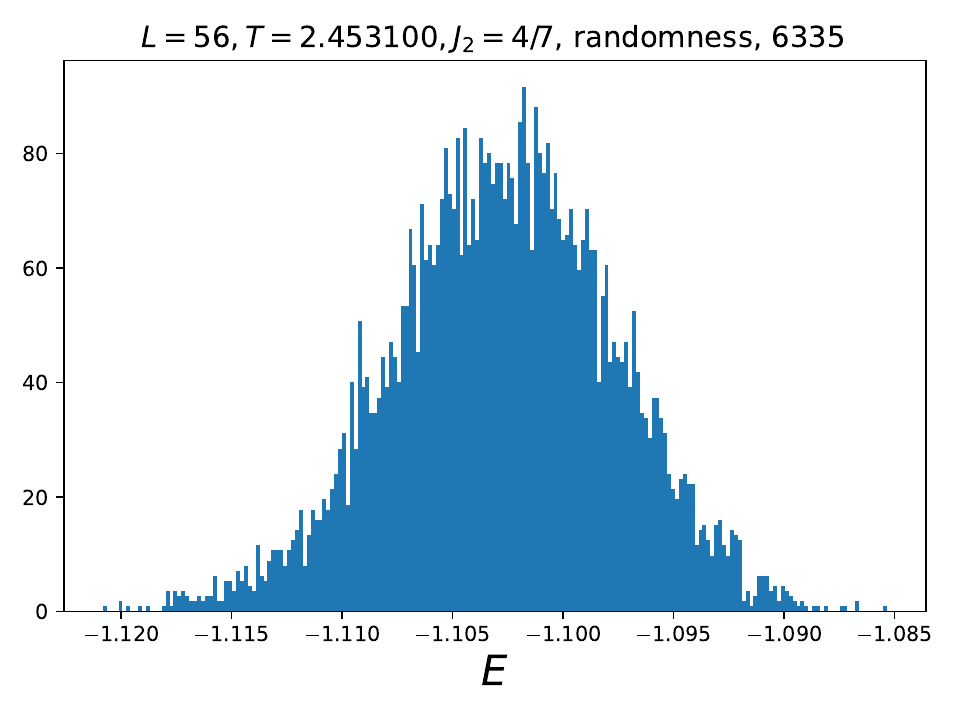}
		}
	\end{center}
	\caption{Histograms of $E$ for $J_2=4/7$ and $L=56$. The initial configurations to start the MC simulations
		and the related number of data points are listed in the titles of the associated panels.
		Data points are recorded once after conducting 2000 MC sweeps.
	}\label{4/7L56}
\end{figure}

For a first-order phase transition in three spatial dimensions, the appearance of energy tunneling, which results in the two-peak structure
in the histograms of $E$, should be suppressed exponentially as $\exp(-c\beta L^{2})$ ($\beta$ is the inverse temperature and $c$ is determined by the interfacial tension).
Interestingly, while MC computations with staggered (middle panel of Fig.~\ref{initial_conf}) and random (right panel of Fig.~\ref{initial_conf}) configurations as the initial configurations require many MC sweeps before one can observe the tunneling phenomena, for simulations using one of the ground-state stripe configurations as the initial configuration, the number of MC sweeps needed is (much) smaller than that for the other two initial configurations;
see Fig.~\ref{4/7L56} for such a scenario. It will be worth exploring this further. In conclusion,
the phase transition of $J_2 = 4/7$ is first order beyond any doubt. 

\begin{figure}
	
	\begin{center}
		\hbox{
			\includegraphics[width=0.33\textwidth]{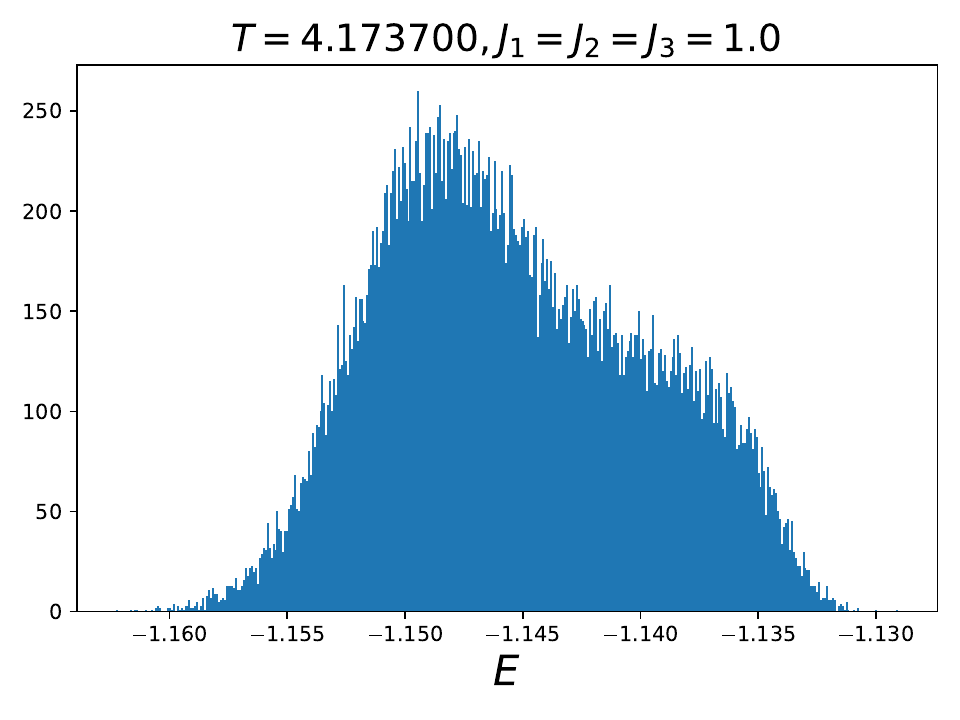}
			\includegraphics[width=0.33\textwidth]{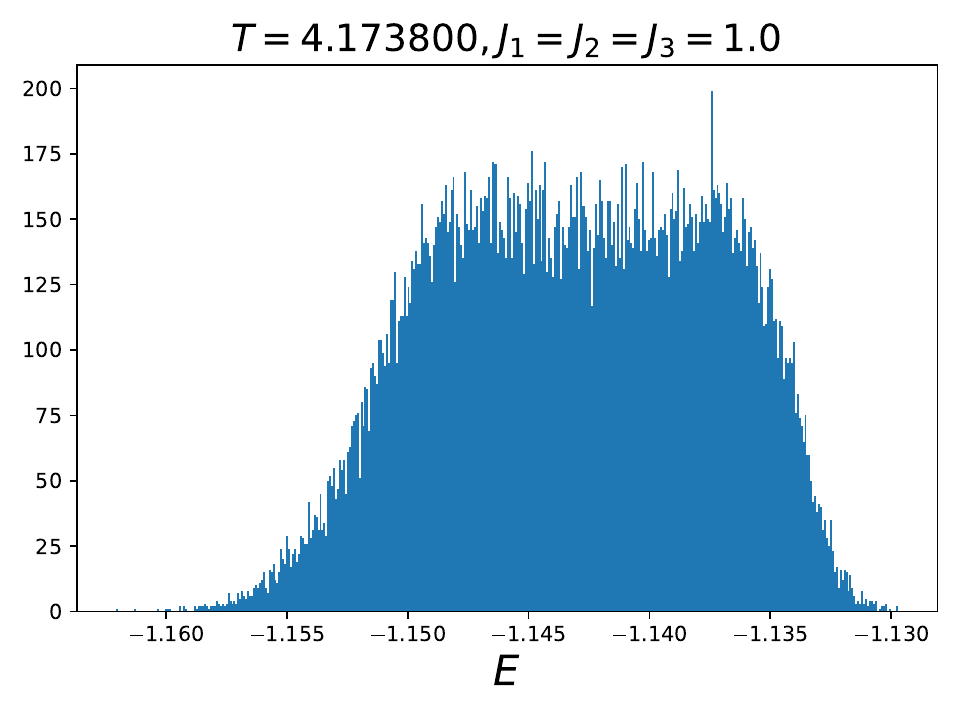}
			\includegraphics[width=0.33\textwidth]{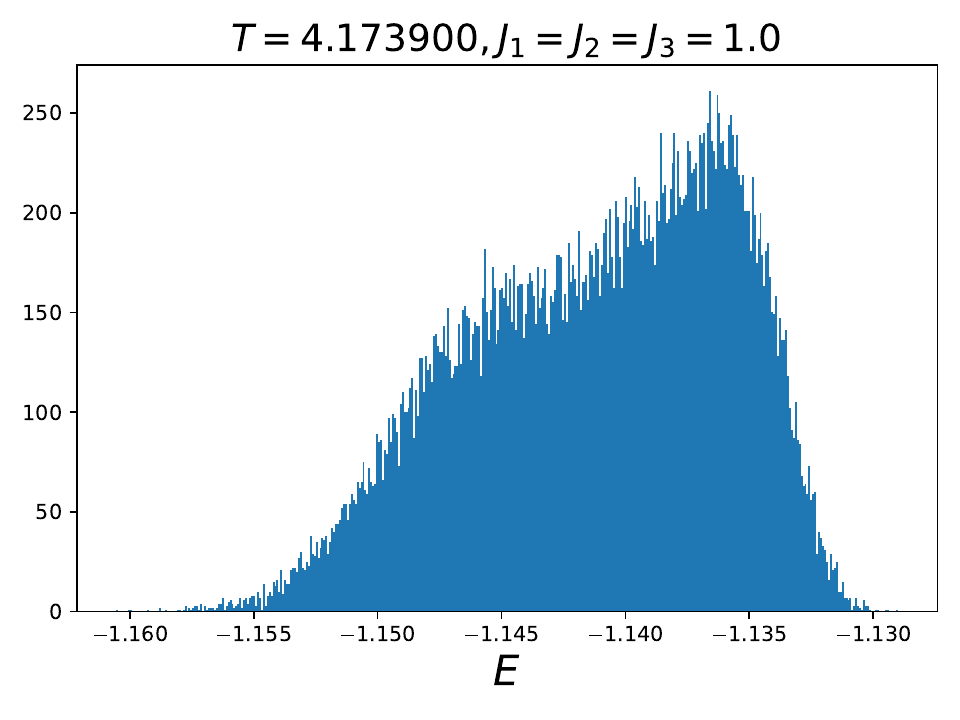}
		}
	\end{center}
	\caption{Histograms of $E$ for $J_2=1.0$ and $L=256$ at three temperatures close to the critical temperature $T_c$. The initial configuration used to start the MC simulations is one of the ground-state stripe configurations.
	}\label{11L256}
\end{figure}

\begin{figure}

	\begin{center}
		\includegraphics[width=0.6\textwidth]{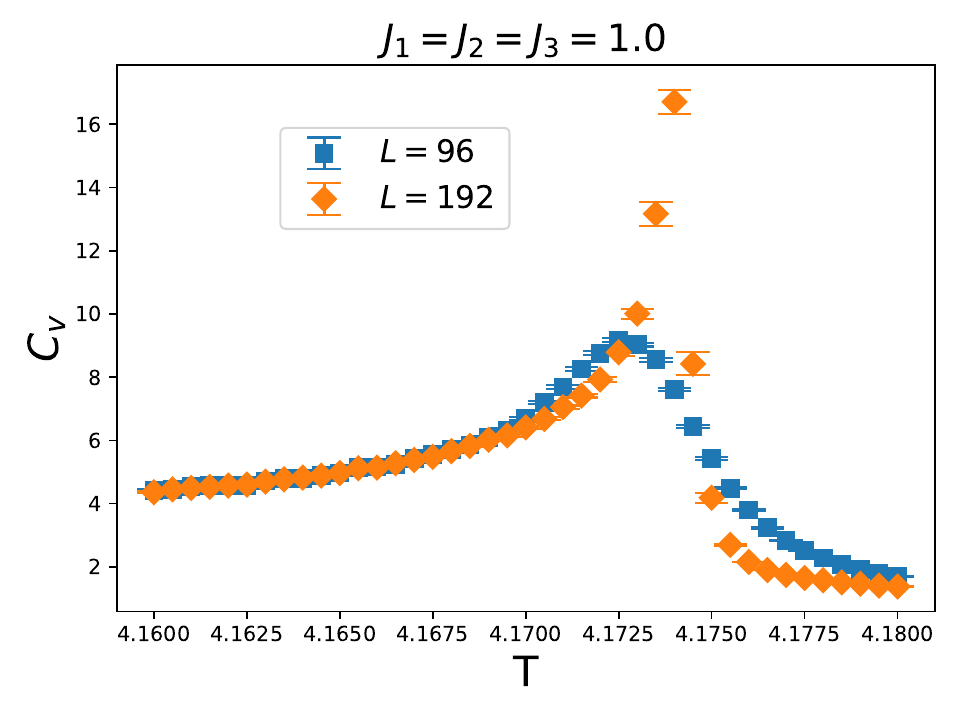}
	\end{center}
	\caption{$C_v$ as functions of $T$ for $J_2 = 1.0$ and $L=96$, $192$.
	}	\label{cvJ11}
\end{figure}

Fig.~\ref{11L256} contains the histograms of $E$ for $J_2=1.0$ and $L=256$ at three temperatures. No obvious sign of a two-peak structure is found in Fig.~\ref{11L256}. In Ref.~\cite{Sor22}, mild two-peak signals are observed for $L = 90$ and $128$. If the phase transition is indeed first order, then the two-peak structure should be more noticeable for $L=256$, which is not the case here. In other words, the data of $J_2 = 1.0$ support the scenario that the phase transition for $J_2 = 1.0$ is second order.
The $T_c$ obtained here for $J_2 = 1.0$
is given by $T_c \sim 4.1738$, which agrees quantitatively with those determined in Refs.~\cite{Ram15,Sor18,Sor22}.
Finally, if the phase transition is first order, then one expects the peak of the specific heat $C_v$ will scale
with $L^3$ in spatial dimension three. In Fig.~\ref{cvJ11}, curves of $C_v$ as functions of $T$ for $L=96$ and $L=192$ are given. Without doubt,
the $L^3$ scaling does not occur in Fig.~\ref{cvJ11}. This provides further evidence that our data favor the scenario that the phase transition
of $J_2 = 1.0$ is second order, not first order.

\subsection{The NN results}

Apart from Monte Carlo investigations, we also study the targeted phase transitions using the NN method.
The considered NN has an extremely simple architecture. Specifically, it is a multi-layer perceptron consisting
of one input layer, one hidden layer, and one output layer. This MLP and its variants have been applied to
explore the critical behaviors of many models. The details of the framework of the MLP used, including the 
activation functions, regularization, and optimizer, can be found in Refs.~\cite{Tan20.1,Tse22,Pen23,Tse241,Tse242}.

\subsubsection{Training and testing sets} 

The considered model is a 3D system. Therefore, using the whole spin configurations in the NN computations for large $L$ will be a challenge. This is
because huge amounts of storage space and memory are required. Here we use a different approach so that
one can easily access the NN results for $L=256$.
Theoretically, the ground-state spin arrangement in each $x$-$y$ layer belongs to one of the four stripe configurations. Moreover, the interlayer nearest-neighbor spins have a strong tendency to be different from each other in the low-temperature region. As a result, the spin arrangement on the $x$-$z$ plane with $y=0$ or the $x$-$y$ plane with $z=0$ of a ground-state configuration takes the form of either one of the four stripe patterns or one of the two staggered (N\'eel) patterns. With this observation, the training set used for training the MLP is made up of these 6
configurations, namely four stripe and two staggered configurations on square lattice. 

\begin{figure}
	
	\begin{center}
			\includegraphics[width=0.6\textwidth]{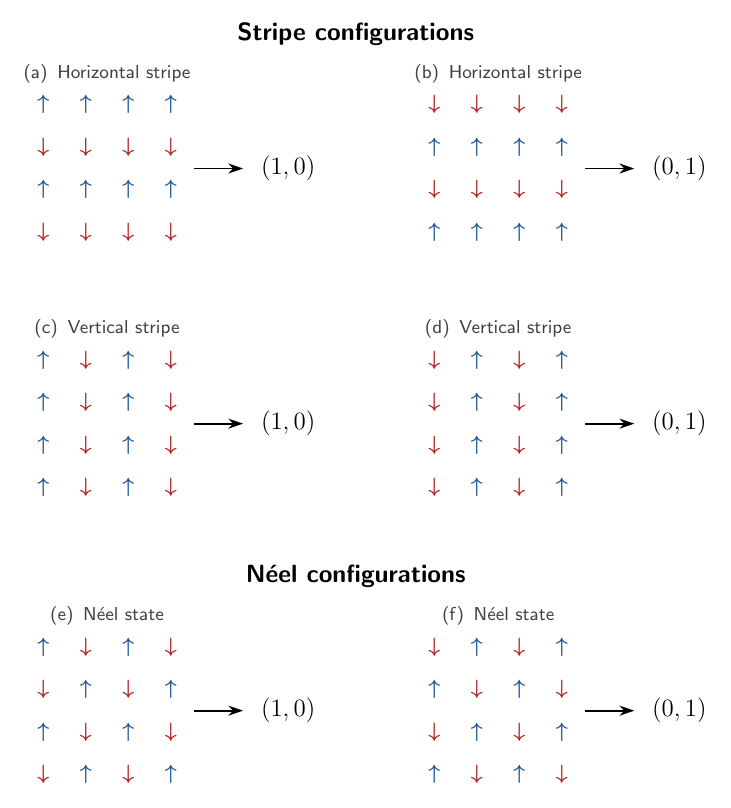}
	\end{center}
	\caption{The assignment of NN output labels to the configurations in the training set. Arrows pointing
		upward (downward) stand for the spins having values of 1 (-1).
	}\label{labels}
\end{figure}

The NN outputs are two-component vectors. Therefore, the employed labels are (1,0) and (0,1). In addition,
a graphical representation of the assignment of NN output labels to the configurations in the training set is given in Fig.~\ref{labels}. With this setup, one expects that a configuration in the low-$T$
region will lead to a label $\sim$ (1,0) or (0,1), which has a magnitude $R = 1$. In addition, for a configuration
in the high-$T$ region ($T > T_c$), one would arrive at a label $\sim$ (0.5,0.5), which has a magnitude 
$R = \frac{1}{\sqrt{2}}$. By investigating the $T$ dependence of $R$, one can estimate the critical temperature $T$. Specifically, for a given $L$, the NN-related pseudo-critical temperature $T_c(L)$ 
is estimated to be the intersection of the $T$-$R$ curve and the horizontal line $0.5(1+1/\sqrt{2})$.
Finally, as we will demonstrate later, the nature of the phase transition can be determined through the probability distribution of $R$.

\subsubsection{$T_c$ computed from NN}

As pointed out previously, for a given $L$, using all $L \times L \times L$ spins as the training set
will limit access to results related to large $L$. Therefore, we conduct the training using 
6 types of $L$ by $L$ artificially made configurations as the training set. Hence, to perform the NN prediction, for an $L^3$ 
configuration generated from the MC simulation, only the subset with $y=0$ (or $z=0$) is considered as one sample in the testing set. 

\begin{figure}
	
	\begin{center}
		\includegraphics[width=0.6\textwidth]{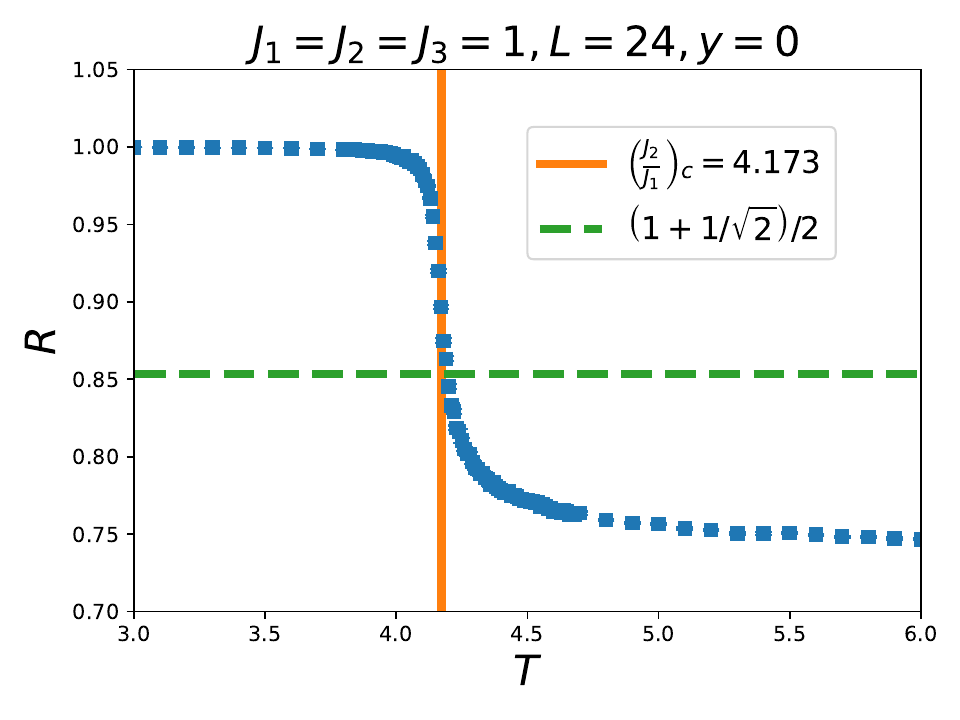}
	\end{center}
	\caption{$R$ as a function of $T$ for $J_2 = 1.0$ and $L=24$. The vertical solid line is the established
		$T_c \sim 4.173$ in the literature and the horizontal dashed line is $0.5(1+1/\sqrt{2})$.
	}\label{J21R}
\end{figure}

First, we will focus on $J_2 = 1.0$ since it is established rigorously that $T_c \sim 4.173$. The $T$ dependence of
$R$ for $J_2 = 1.0$ and $L = 24$ is given in Fig.~\ref{J21R}. In the figure, the horizontal dashed and vertical solid lines are $0.5(1+1/\sqrt{2})$ and the expected $T_c \sim 4.173$, respectively. As can be seen from the
figure, the intersection of the horizontal dashed line with the $T$-$R$ curve matches nicely with the vertical solid line, which is the expected $T_c$. This result provides convincing evidence that our approach, namely 
employing only the subset with $y=0$, which is effectively a 2D system, as the testing set, can accurately locate the $T_c$ of the 3D system.

\begin{figure}
	
	\begin{center}
		\includegraphics[width=0.6\textwidth]{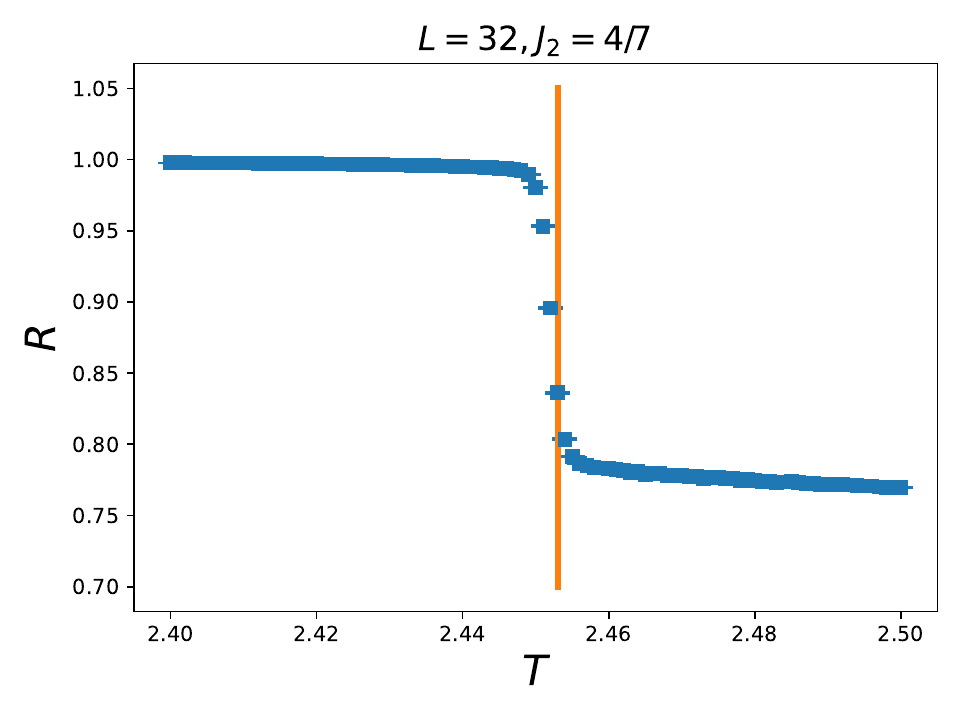}
	\end{center}
	\caption{$R$ as a function of $T$ for $J_2 = 4/7$, $L=32$ and $y=0$. The vertical solid line is the $T_c$ estimated from the MC data.
	}\label{J247R}
\end{figure}

$R$ as a function of $T$ for $J_2 = 4/7$ and $L=32$ is shown in Fig.~\ref{J247R}. The vertical line in the 
figure is the $T_c$ estimated from the MC data. As can be seen from the figure, $R$ turns sharply from
1 to (around) $1/\sqrt{2}$ close to the vertical line. This again confirms the validity of our NN computation approach. Specifically, with the training strategy, even using only the subset of $y=0$, one can determine the critical
temperature precisely. Such a method dramatically reduces the storage space and computing resources
needed to carry out the NN investigation for large lattices of a 3D system.

\subsubsection{Nature of the studied phase transitions}

\begin{figure}
	
	\begin{center}
		\hbox{~~~~~~~~~~
			\includegraphics[width=0.4\textwidth]{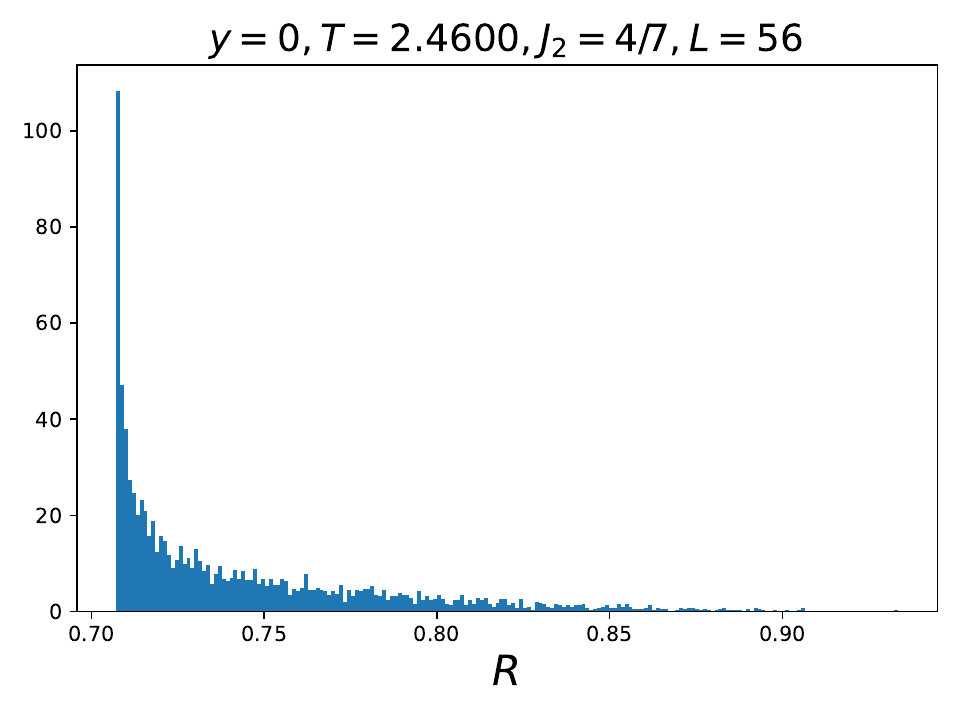}~~~~~~~
			\includegraphics[width=0.4\textwidth]{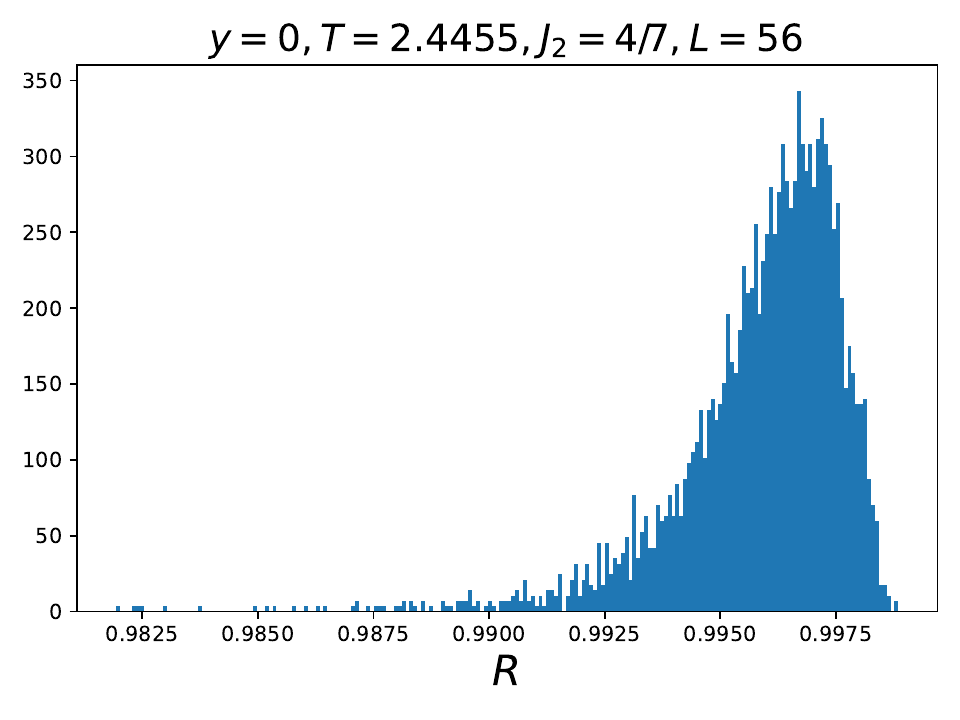}
		}
	\end{center}
	\caption{Histograms of $R$ for $J_2 = 4/7$, $y=0$, and $L=56$ for $T=2.460$ (left) and $T = 2.4455$ (right).
	}\label{RJ47far}
\end{figure}

\begin{figure}
	
	\begin{center}
		\includegraphics[width=0.6\textwidth]{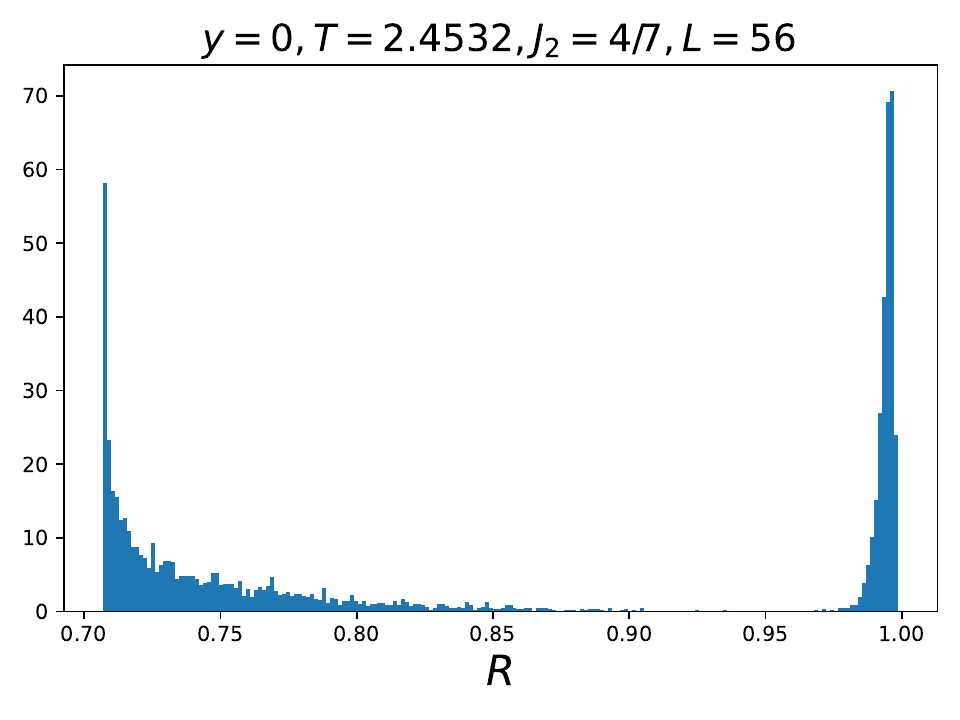}
	\end{center}
	\caption{Histograms of $R$ for $J_2 = 4/7$, $y = 0$, and $L=56$ at $T=2.4532$.
	}\label{RJ47Tc}
\end{figure}

\begin{figure}
	
	\begin{center}
		\includegraphics[width=0.6\textwidth]{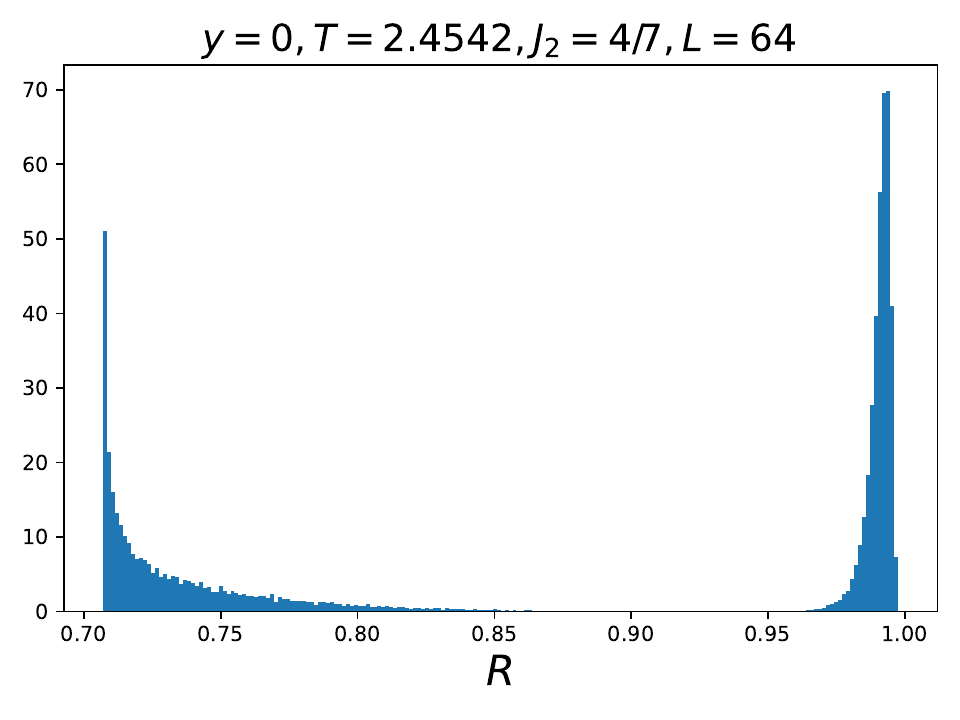}
	\end{center}
	\caption{Histograms of $R$ for $J_2 = 4/7$, $y=0$, and $L=64$ at $T=2.4542$.
	}\label{RJ4764}
\end{figure}

\begin{figure}
	
	\begin{center}
		\vbox{
			\hbox{~~~~~~~~
			\includegraphics[width=0.3\textwidth]{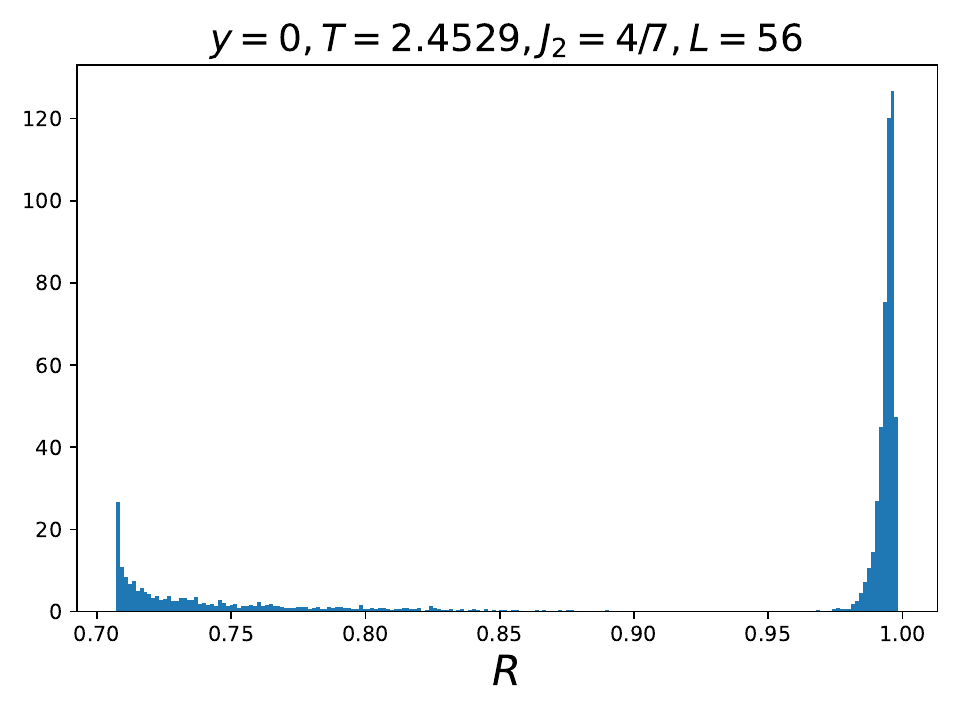}
			\includegraphics[width=0.3\textwidth]{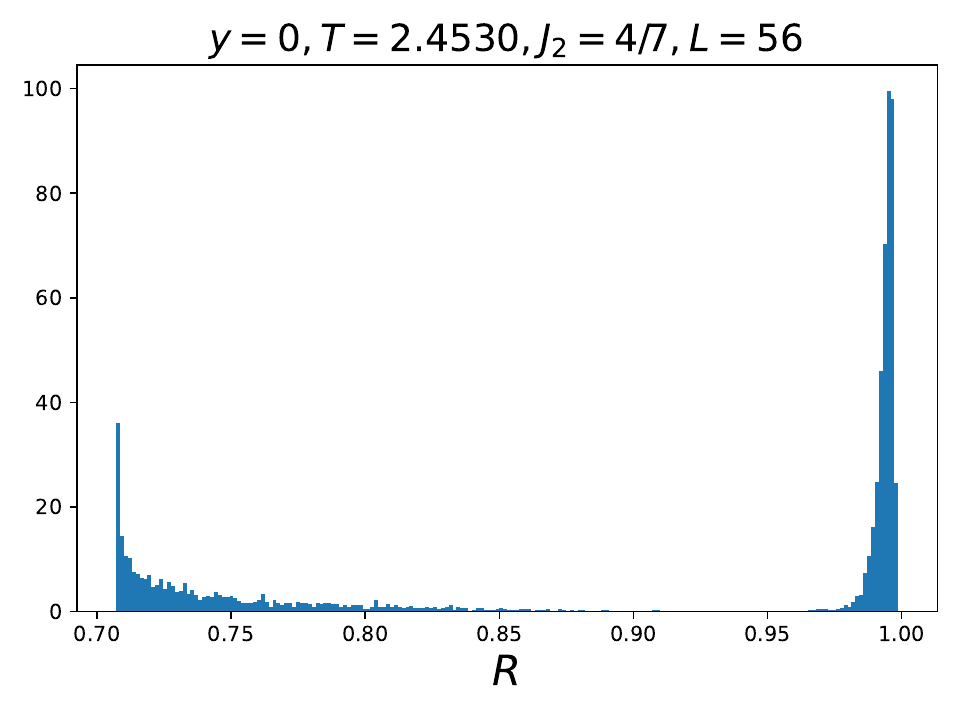}
			\includegraphics[width=0.3\textwidth]{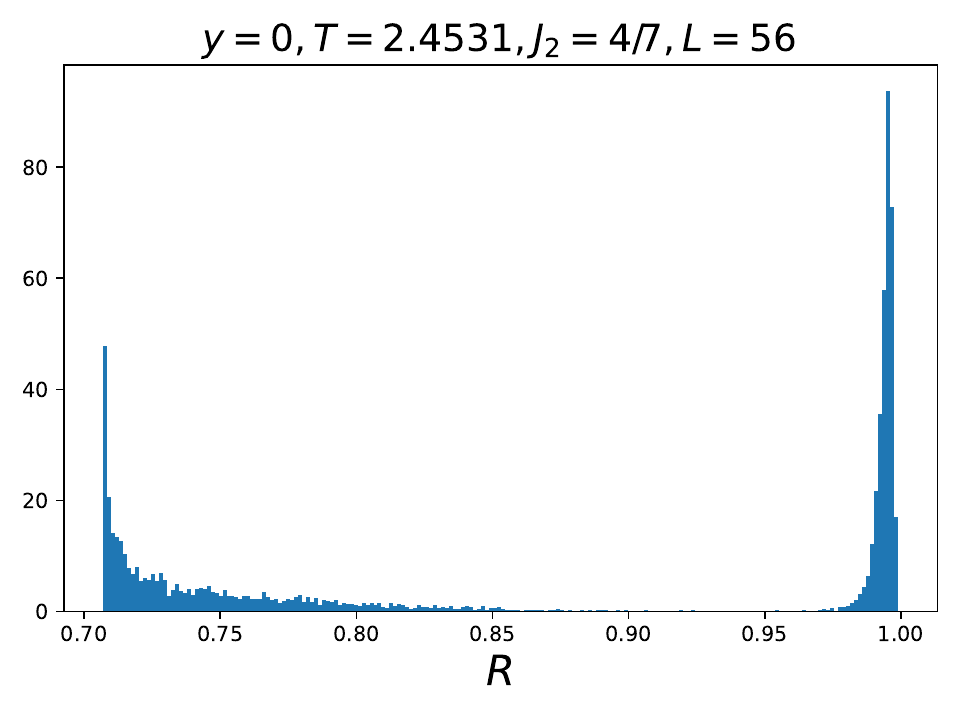}
		}
		\hbox{~~~~~~~~~	\includegraphics[width=0.3\textwidth]{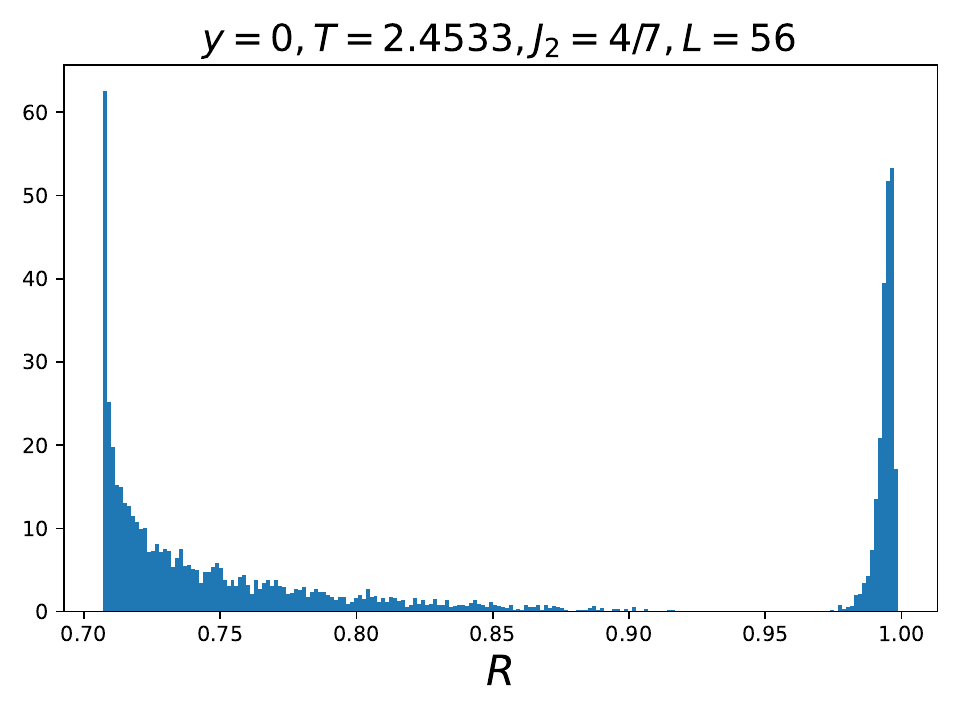}
			\includegraphics[width=0.3\textwidth]{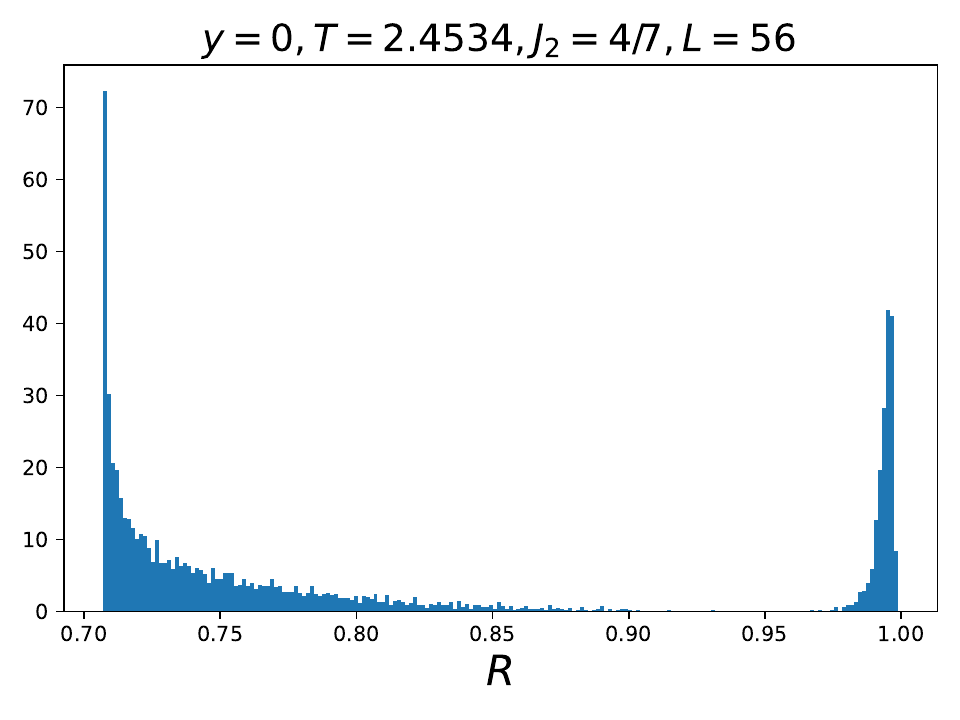}
			\includegraphics[width=0.3\textwidth]{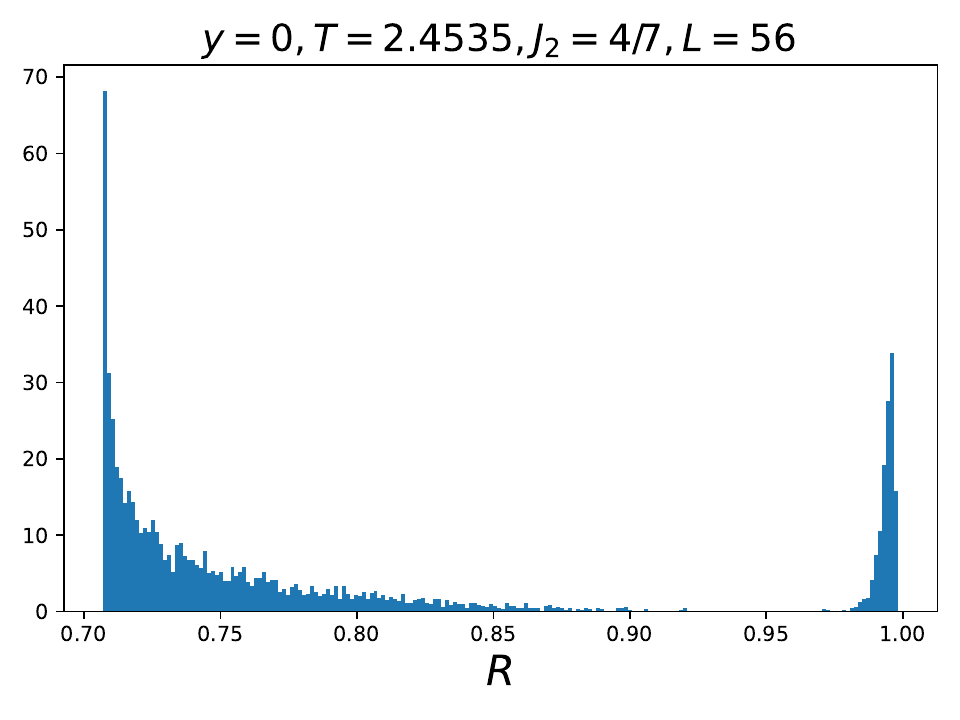}
	}
		}
	\end{center}
	\caption{Histograms of $R$ for $J_2 = 4/7$, $y=0$ and $L=56$ for several $T$ below (top) and above (bottom) $T_c$.
	}\label{RJ47yabovebelow}
\end{figure}

The histograms of $R$ for $J_2 = 4/7$ and $L=56$ for $T=2.460$ and $T = 2.4455$ are shown in the left and the right panels of Fig.~\ref{RJ47far}, respectively. The temperatures $T=0.2460$ and $T=2.4455$ are above and below the $T_c$ obtained from MC data, respectively. The results of the figure show that the values of $R$ for $T < T_c$ are located in a narrow region close to 1. In addition, the majority of the values of $R$ for $T > T_c$ are near $1/\sqrt{2}$
and a small number of values of $R$ spread from $1/\sqrt{2}$ to a value moderately below 1. In particular, only very few
values of $R$ (or none) can reach the region close to 1. The histogram of $R$ for $J_2 = 4/7$ and $L=56$ 
at $T = 2.4532$ is depicted in Fig.~\ref{RJ47Tc}. $T=2.4532$ is close to the estimate of $T_c$ from MC data.
In addition, the distribution of $R$ shown in Fig.~\ref{RJ47Tc} has features of both the left and the right
panels of Fig.~\ref{RJ47far}. This implies the $R$ data at $T=2.4532$ contain those associated with both $T < T_c$ and $T > T_c$, which will lead to the conclusion that the phase transition of $J_2 = 4/7$ is first order.
Finally, Fig.~\ref{RJ4764} shows the histogram of $R$ for $J_2 = 4/7$ and $L=64$ at $T = 2.4542 \sim T_c$.
The separation of the distribution of $R$ into two peaks at 1 and $1/\sqrt{2}$ is more obvious than that for $L=56$. With all the findings demonstrated above, one arrives at the conclusion that the phase transition
of $J_2 = 4/7$ is first order without ambiguity.   
It is also interesting to investigate the histograms of $R$ close to $T_c$. Fig.~\ref{RJ47yabovebelow} shows the histograms of $R$ for $L=56$ and $T = 2.5429, 2.5430, 2.5431, 2.5433, 2.5434, 2.5435$. Clearly, the two-peak structure in $R$, i.e., the signal of coexistence of two phases, appears for those 
$T$ slightly below and above $T_c$ as well.

The results presented above are obtained using the subset of spin configurations with their $y$-coordinate being 0. Since the ground-state configurations in any $x$-$y$ plane belong to any one of the four stripe patterns, with the training strategy employed here, the subset of spin configurations with their $z$-coordinate equal to 0 can be considered to study the targeted phase transition(s) as well.

\begin{figure}
	
	\begin{center}
		\vbox{
			\hbox{~~~~~~~~
				\includegraphics[width=0.3\textwidth]{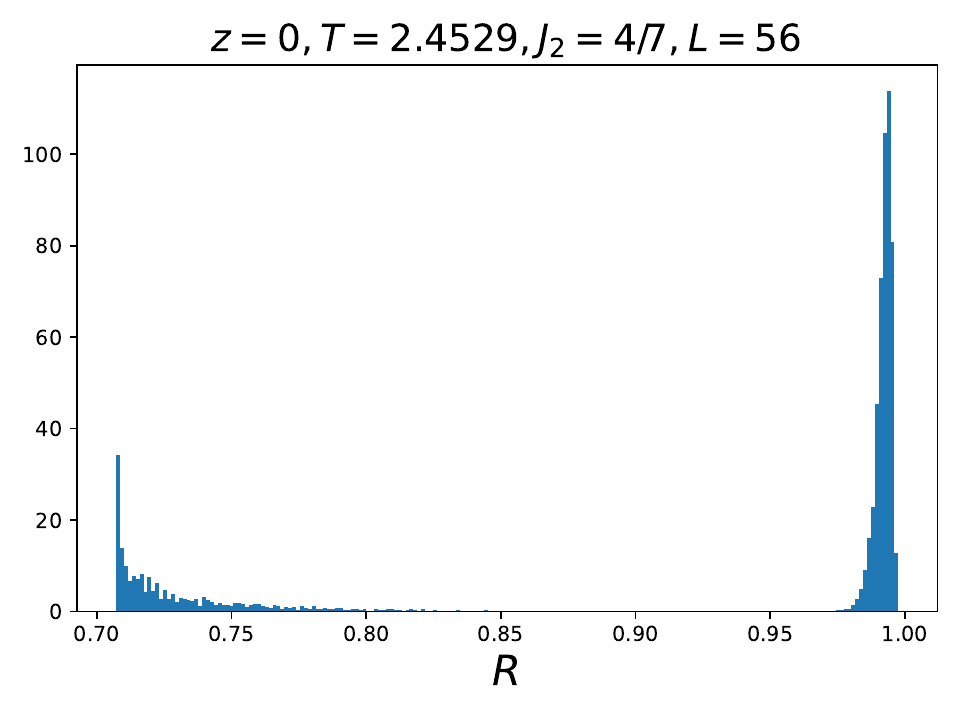}
				\includegraphics[width=0.3\textwidth]{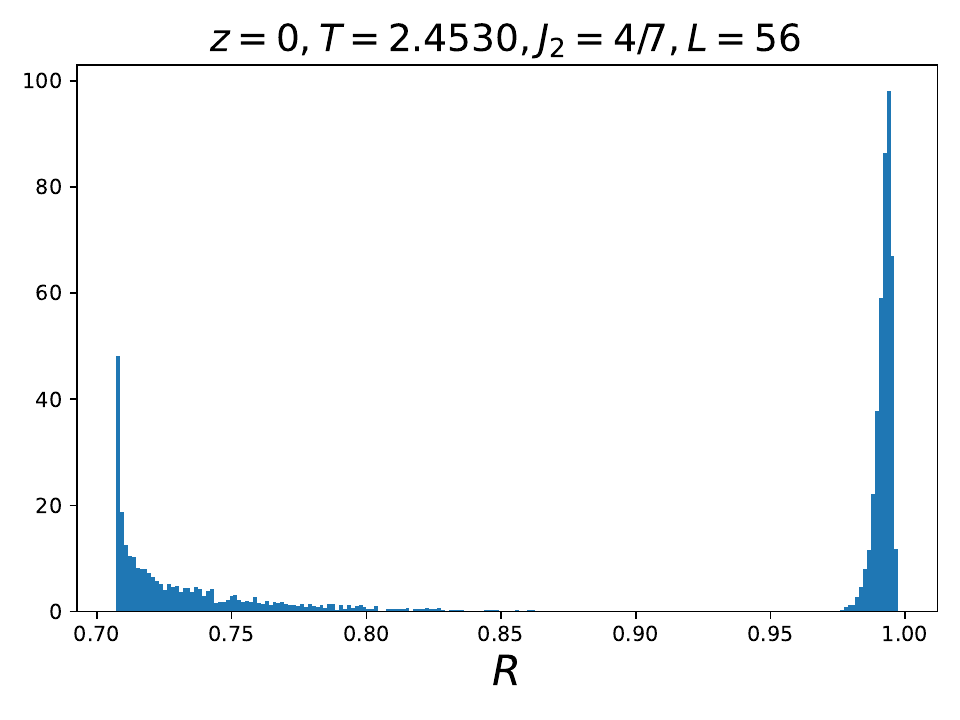}
				\includegraphics[width=0.3\textwidth]{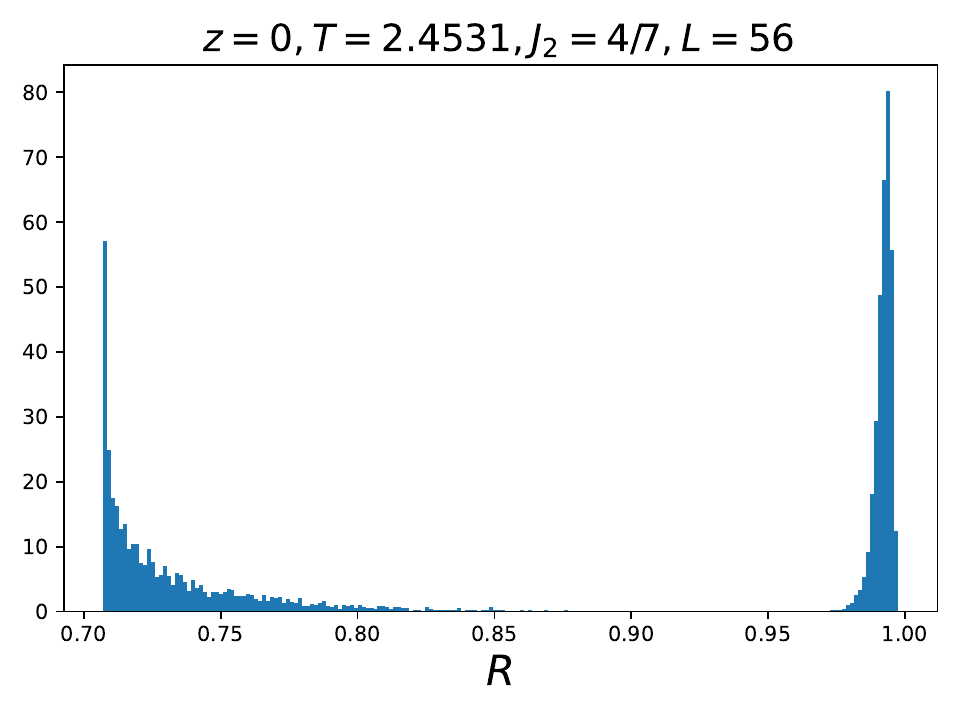}
			}
			\hbox{~~~~~~~~~	\includegraphics[width=0.3\textwidth]{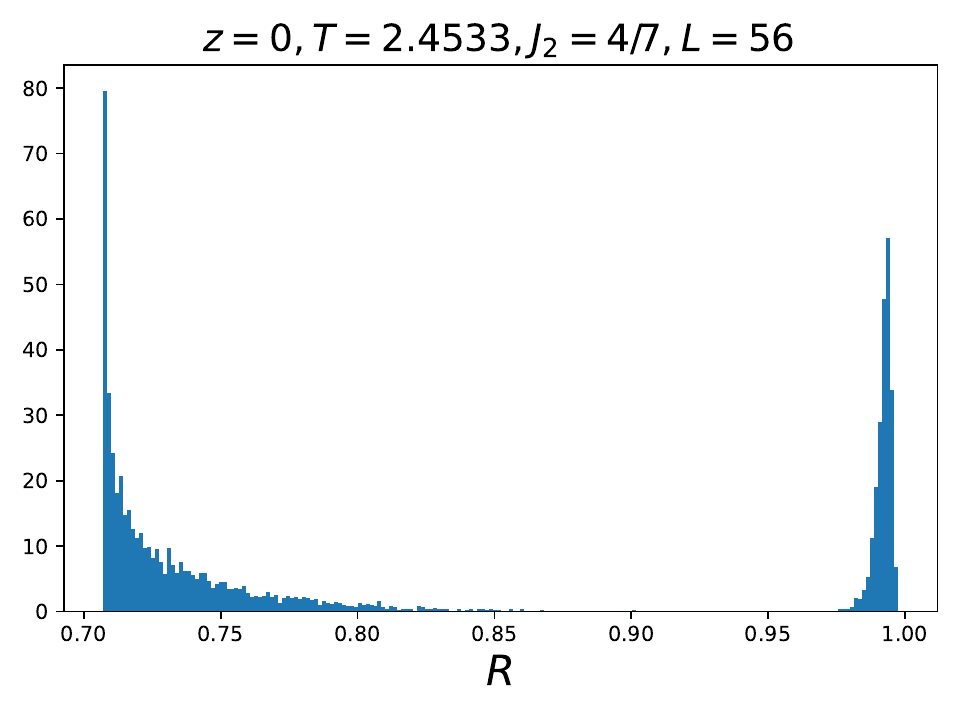}
				\includegraphics[width=0.3\textwidth]{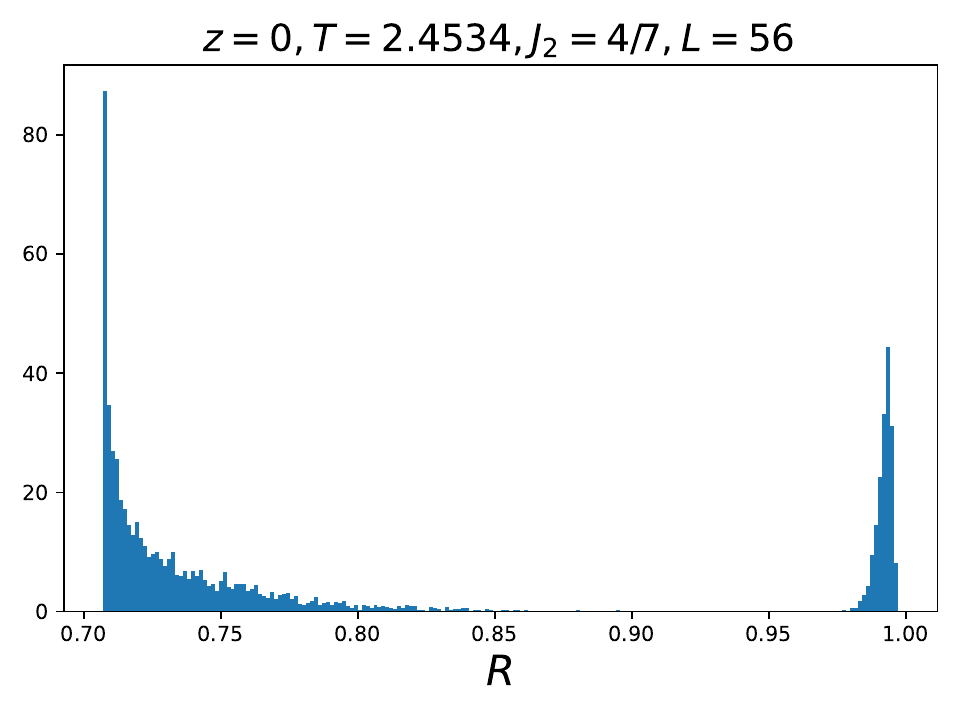}
				\includegraphics[width=0.3\textwidth]{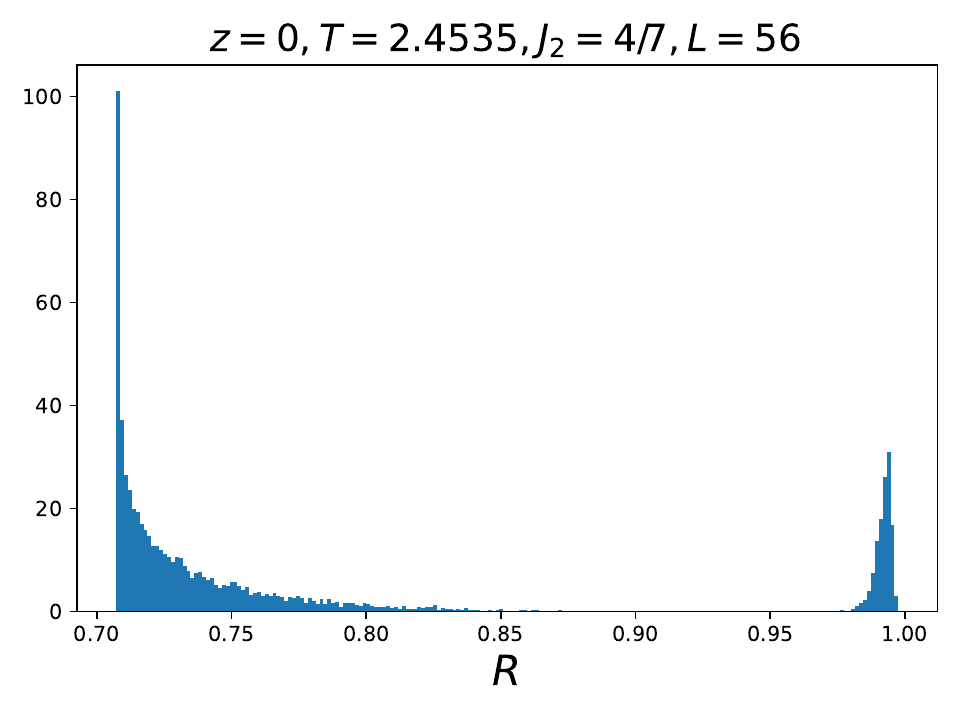}
			}
		}
	\end{center}
	\caption{Histograms of $R$ for $J_2 = 4/7$, $z=0$ and $L=56$ for several $T$ below and above $T_c$.
	}\label{RJ47zabovebelow}
\end{figure}

The histograms of $R$ for $J_2 = 4/7$ and $L=56$ for several $T$ close to $T_c$ are depicted in Fig.~\ref{RJ47zabovebelow}. The associated testing set consists of spin configurations with their $z$-coordinate being 0. The similarity between the histograms shown in Figs.~\ref{RJ47yabovebelow} and
\ref{RJ47zabovebelow} convincingly suggests that both testing sets, namely the configurations with $y=0$ or $z=0$, can be used to explore the critical behaviors of the studied phase transitions.

\begin{figure}
	
	\begin{center}
			\hbox{~~~~~~~~
				\includegraphics[width=0.3\textwidth]{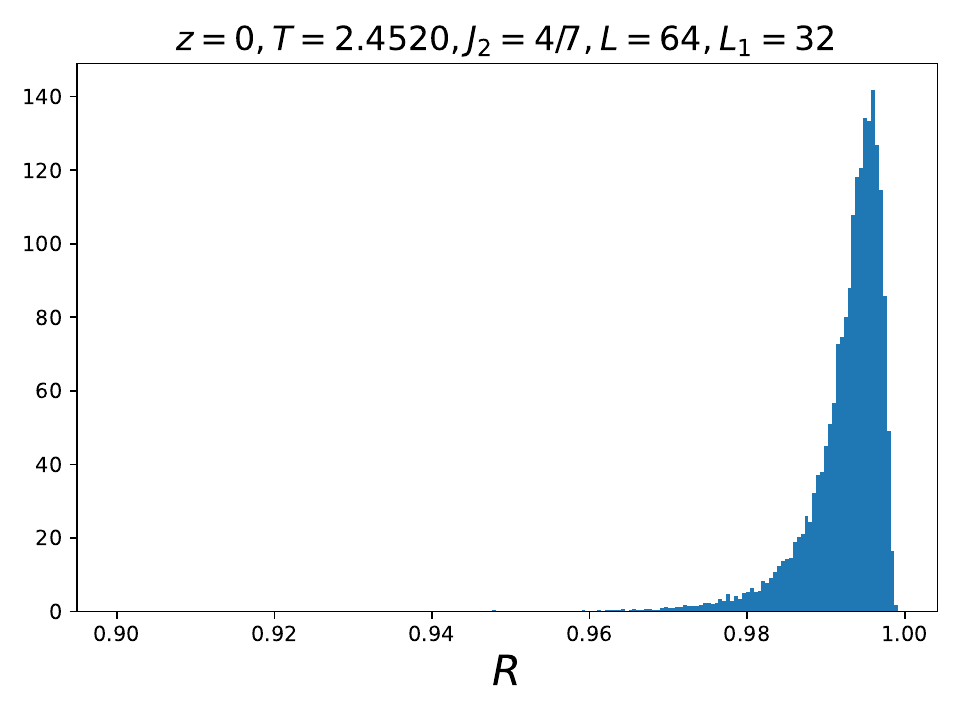}
				\includegraphics[width=0.3\textwidth]{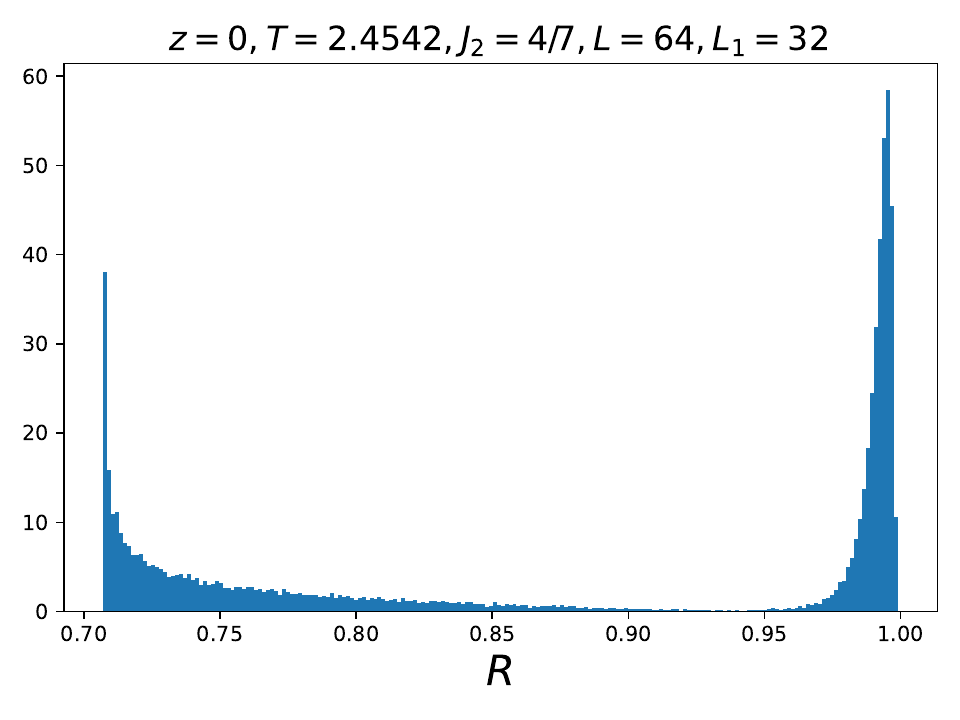}
				\includegraphics[width=0.3\textwidth]{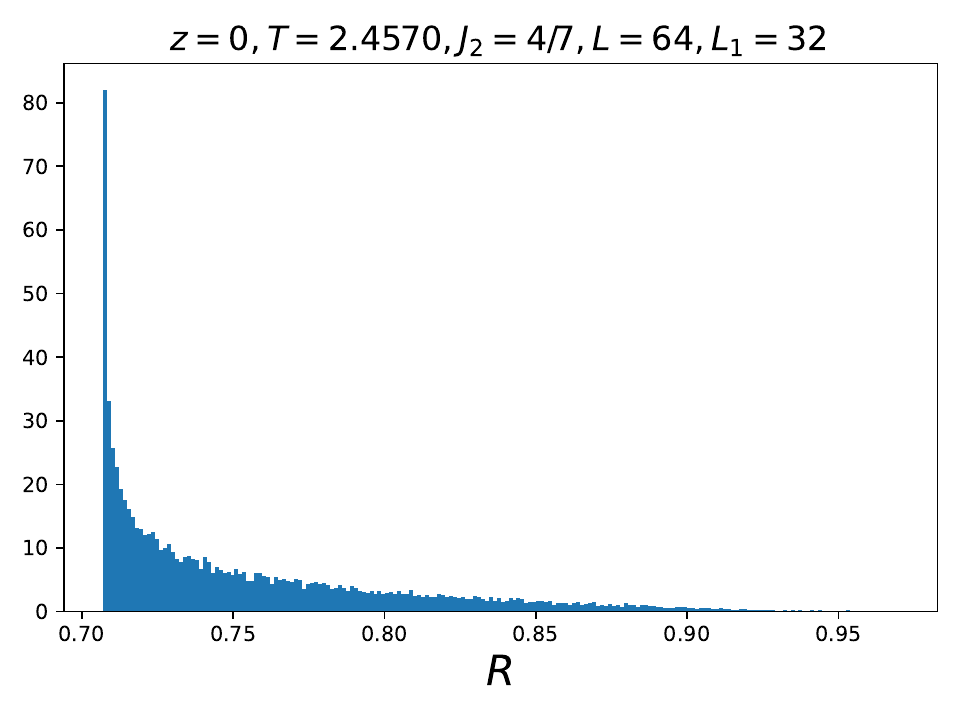}
			}
	\end{center}
	\caption{Histograms of $R$ for $J_2 = 4/7$, $z=0$, $L=64$ and $L_1 = 32$ for $T$ below (left), at (middle), and above (right) the pseudo-critical $T_c(L)$.
	}\label{RJ47zL64L32abovebelow}
\end{figure}

Interestingly, if the bottom left $32 \times 32$ subsets of the whole $64 \times 64$ configurations are considered as the testing set, then the resulting $R$ can be employed to determine the nature of the phase transition as well. The left,
middle, and right panels of Fig.~\ref{RJ47zL64L32abovebelow} demonstrate the histograms of $R$ far below, close to, and far above the associated pseudo-critical temperature $T_c(L)$. The left and right panels reveal the information that
values of $R$ are concentrated around $1$ and $1/\sqrt{2}$, respectively. In addition, the values of $R$ in the middle panel of Fig.~\ref{RJ47zL64L32abovebelow} are distributed mainly around both $1$ and $1/\sqrt{2}$ and hence exhibit features of $R$ in the ordered and disordered phases. This phenomenon can be interpreted as the coexistence of phases below and above the
critical point. In conclusion, even $32 \times 32$ subsets of $64 \times 64 \times 64$ configurations can provide
strong evidence that the phase transition of $J_2 = 4/7$ is first order. Later, we will show that the strategy of using $L_1 \times L_1$ subsets of $L \times L \times L$ configurations as the testing set is also applicable for $J_2 = 1.0$ and $L=256$. These outcomes suggest that our approach
can dramatically reduce the storage capacity and computing resource requirements when conducting NN computations for higher-dimensional systems.

\begin{figure}
	\begin{center}
			\includegraphics[width=0.5\textwidth]{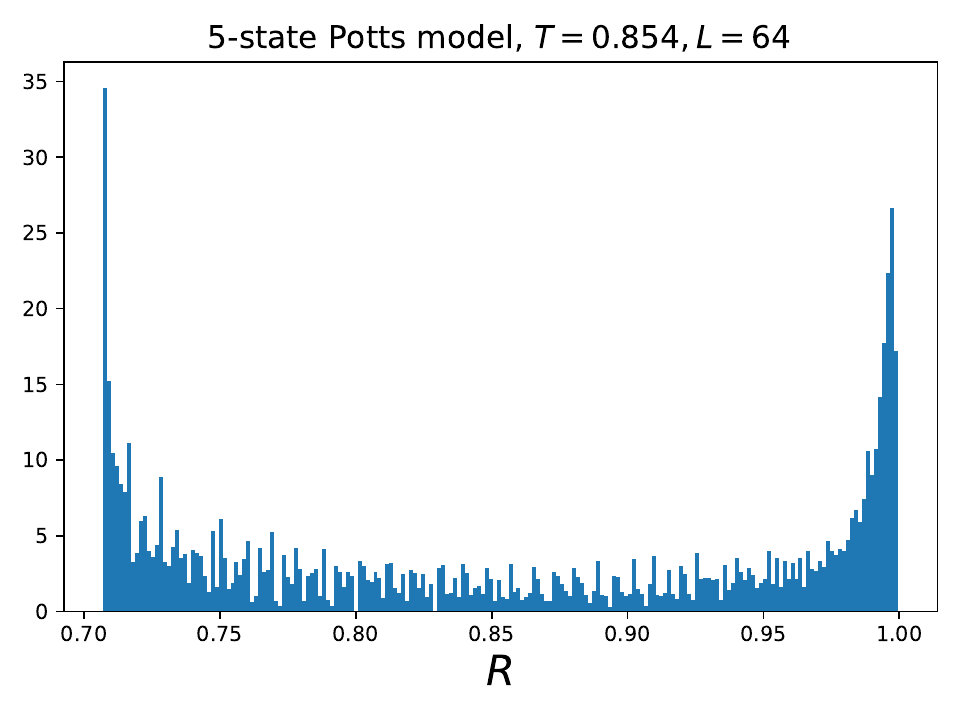}
	\end{center}
	\caption{Histogram of $R$ for the 2D ferromagnetic 5-state Potts model on the square lattice close to the related $T_c$ with $L=64$.
	}\label{potts5}
\end{figure}

Before presenting the results of $R$ corresponding to $J_2 = 1.0$, we would like to point out that for a first-order phase transition and when either $L \times L$ or the subsets $L_1 \times L_1$ are considered as the testing sets, the dominated peaks of the histograms of $R$ close to $T_c$ are located near $1$ and $1/\sqrt{2}$. This is indeed the case for the considered model as can be observed from Figs.~\ref{RJ47Tc}, \ref{RJ4764}, \ref{RJ47yabovebelow}, \ref{RJ47zabovebelow}, and \ref{RJ47zL64L32abovebelow}. The scenario is also valid even for a weak first-order phase transition; see Fig.~\ref{potts5} which is the histogram of $R$ for the 2D ferromagnetic 5-state Potts model on the square lattice close to the corresponding $T_c$ with $L=64$.

\begin{figure}
	
	\begin{center}
		\hbox{~~~~~~~~~~
			\includegraphics[width=0.4\textwidth]{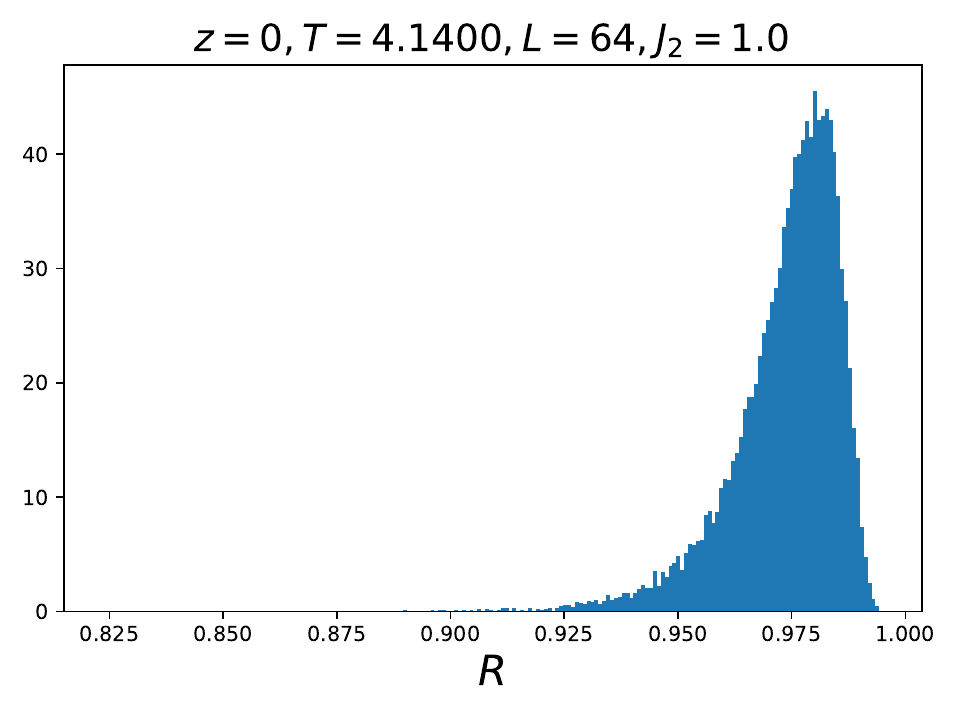}~~~~~~~
			\includegraphics[width=0.4\textwidth]{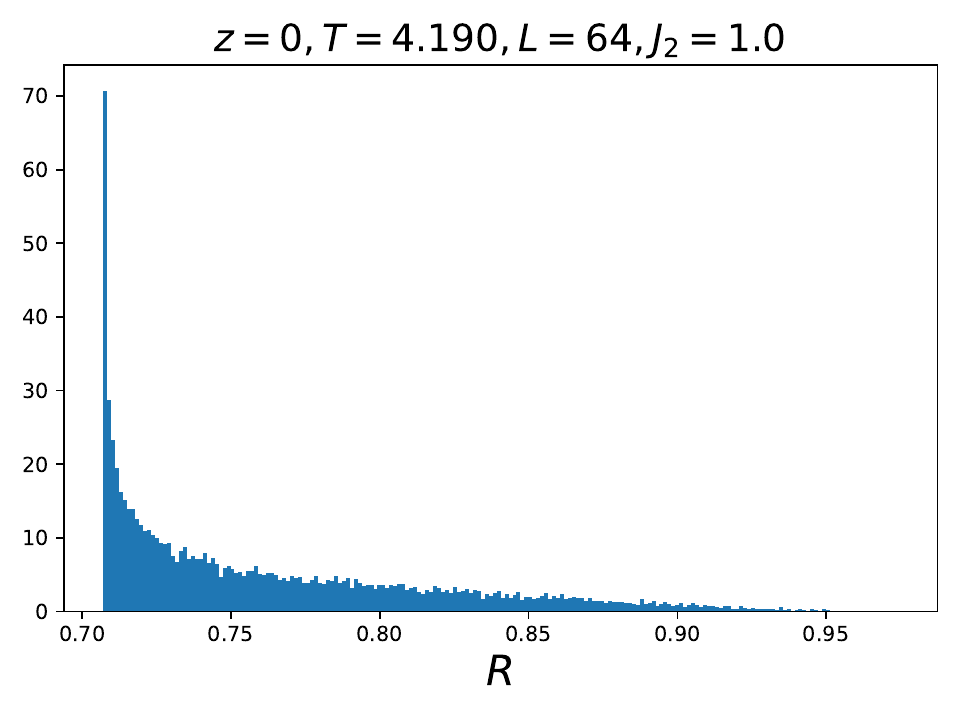}
		}
	\end{center}
	\caption{Histograms of $R$ for $J_2 = 1.0$, $z=0$, and $L=64$ for $T=4.140$ (left) and $T = 4.190$ (right).
	}\label{RJ11far}
\end{figure}

The left and right panels of Fig.~\ref{RJ11far} demonstrate the histograms of $R$ for $J_2 = 1.0$ and $L=64$
at $T=4.140$ and $T=4.190$, respectively. The results are obtained using a testing set consisting of configurations with their $z$-coordinate equal to 0. The outcomes in Fig.~\ref{RJ11far} show the expected behaviors. Specifically, the majority of the values of $R$ far below and above $T_c$ would be located close to 1 and $1/\sqrt{2}$, respectively. This scenario is the same as that of $J_2 = 4/7$. 

\begin{figure}
	
	\begin{center}
		\vbox{
			\hbox{~~~~~~~~
				\includegraphics[width=0.3\textwidth]{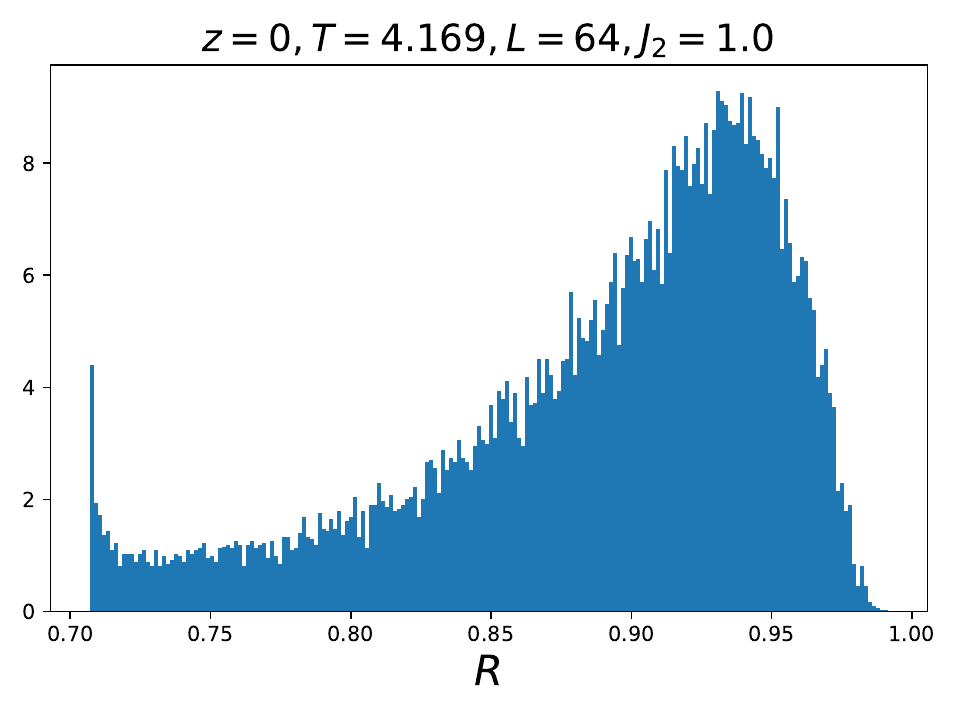}
				\includegraphics[width=0.3\textwidth]{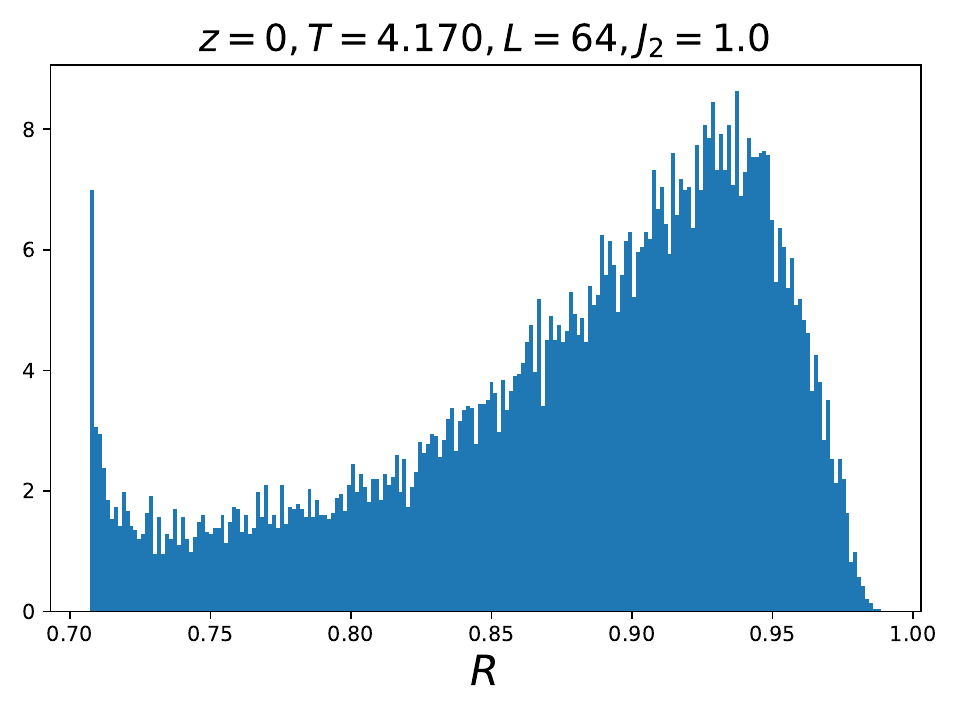}
				\includegraphics[width=0.3\textwidth]{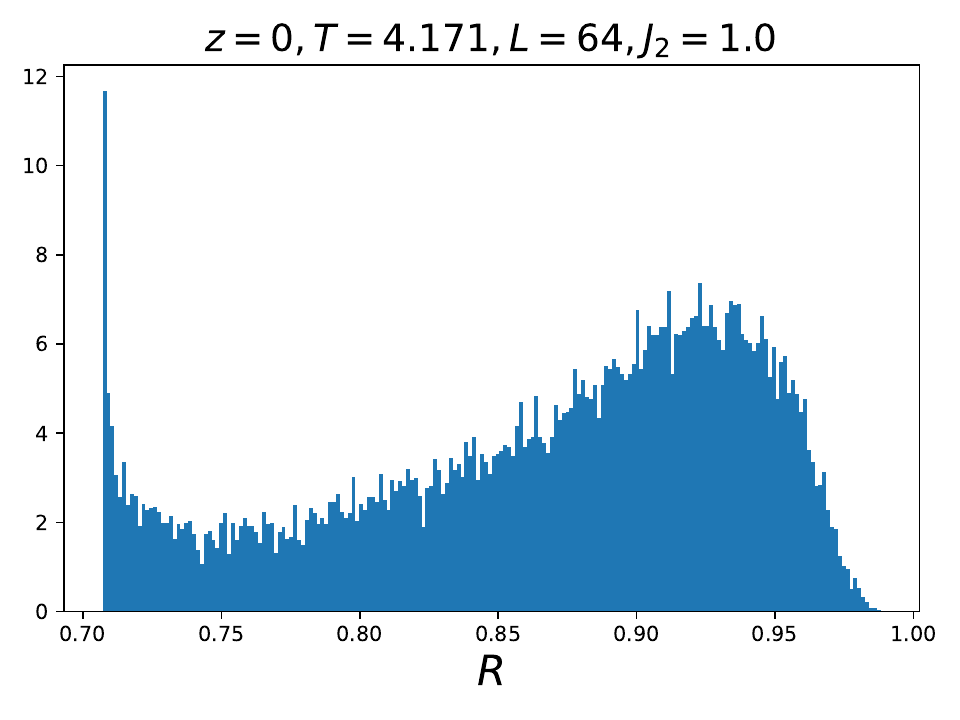}
			}
			\hbox{~~~~~~~~~	\includegraphics[width=0.3\textwidth]{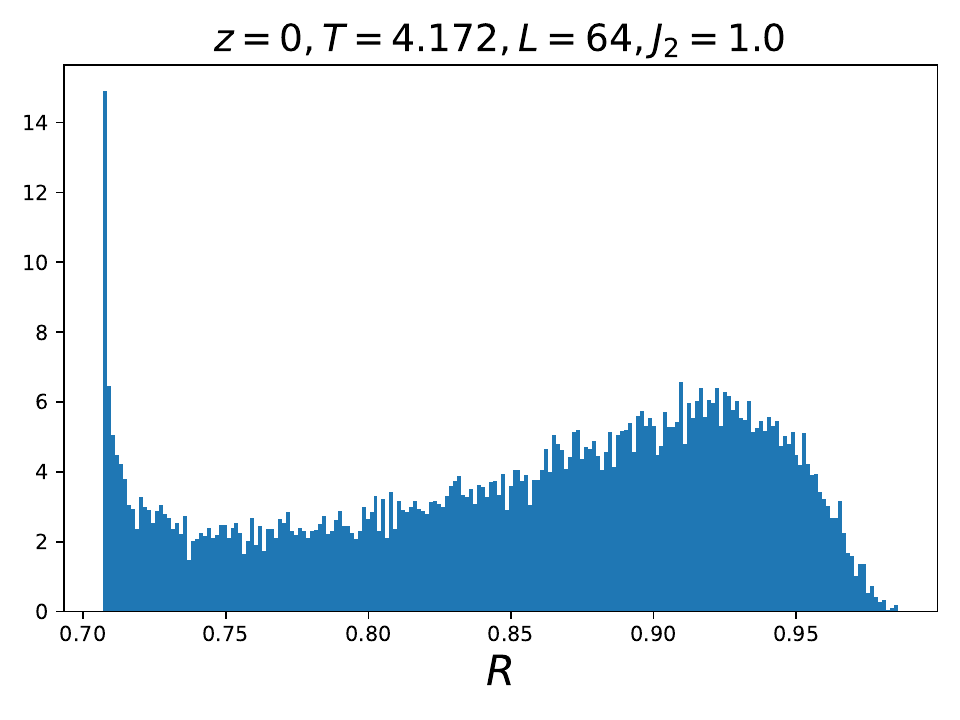}
				\includegraphics[width=0.3\textwidth]{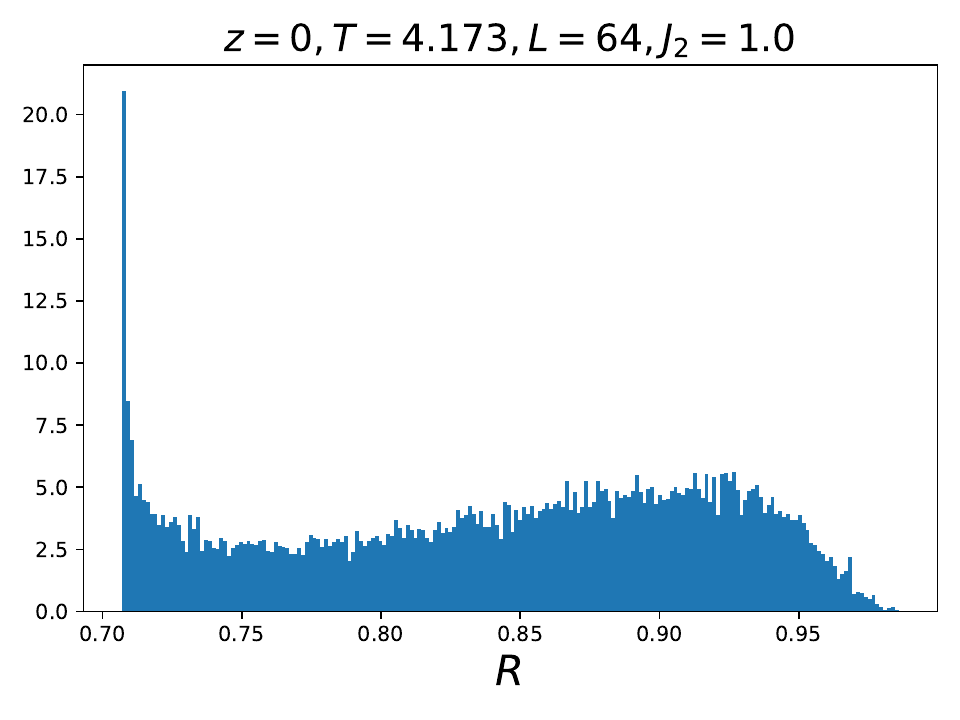}
				\includegraphics[width=0.3\textwidth]{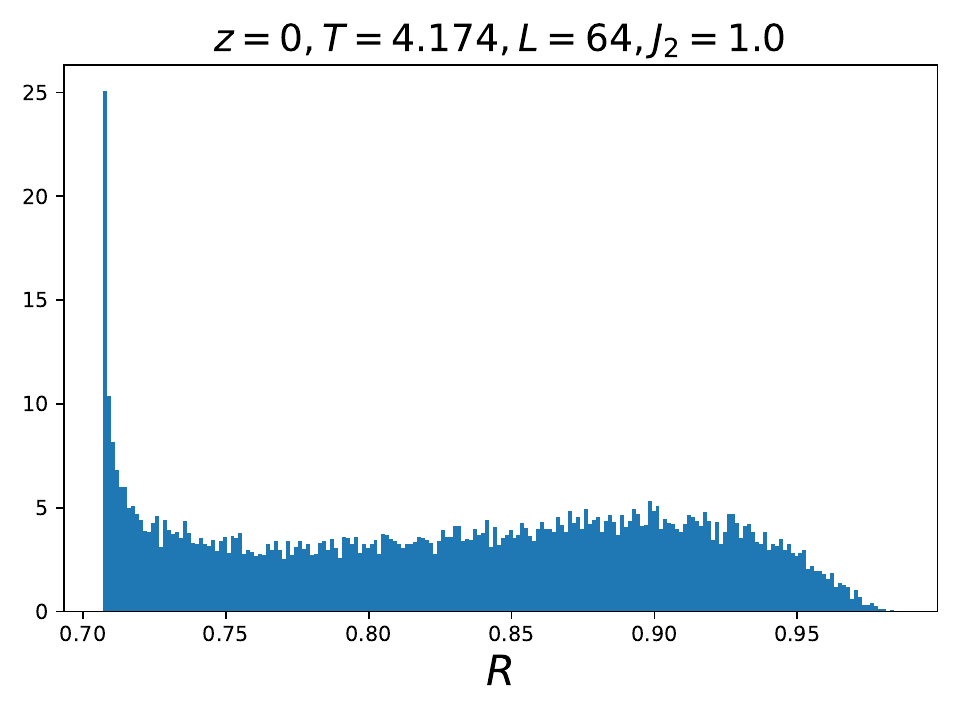}
			}
		}
	\end{center}
	\caption{Histograms of $R$ for $J_2 = 1.0$, $z=0$ and $L=64$ for several $T$ below and above $T_c$.
	}\label{RJ11zabovebelow}
\end{figure}

\begin{figure}
	
	\begin{center}
		\vbox{
			\hbox{~~~~~~~~
				\includegraphics[width=0.3\textwidth]{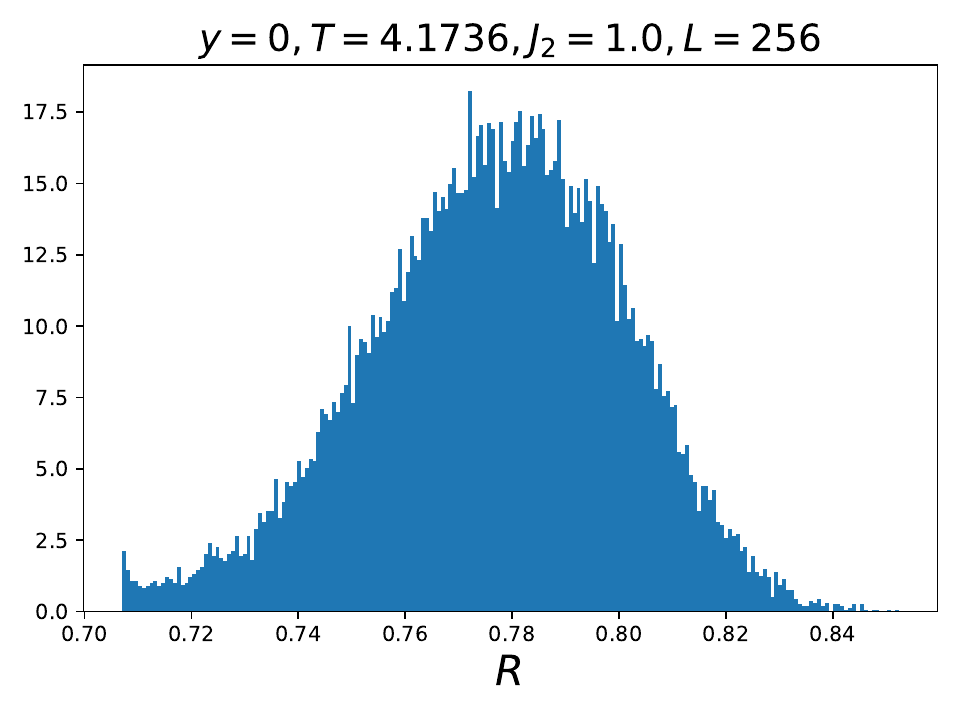}
				\includegraphics[width=0.3\textwidth]{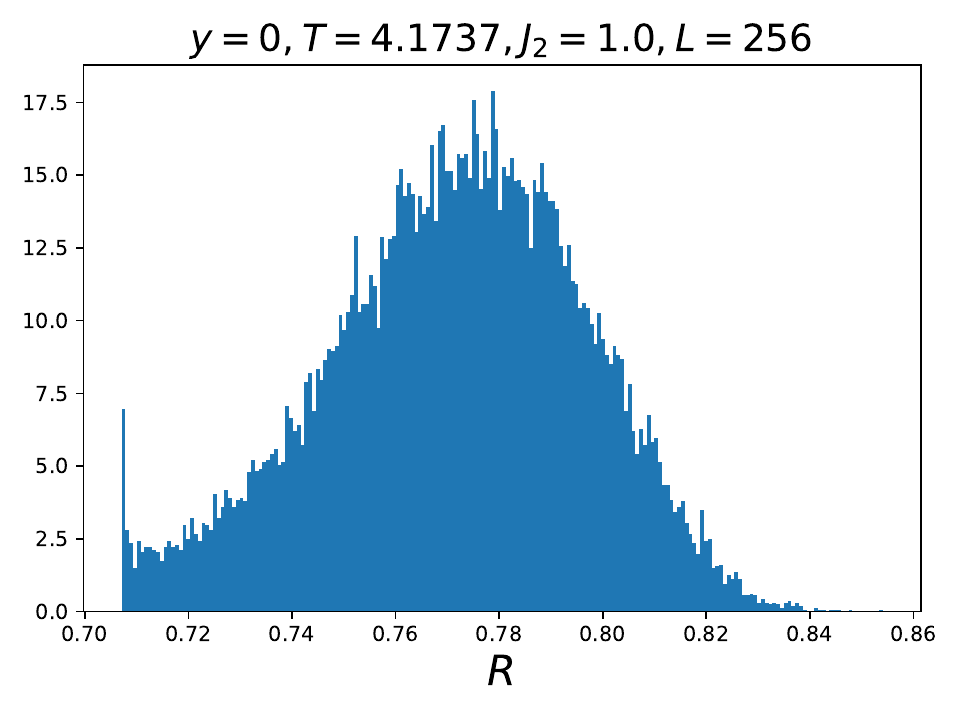}
				\includegraphics[width=0.3\textwidth]{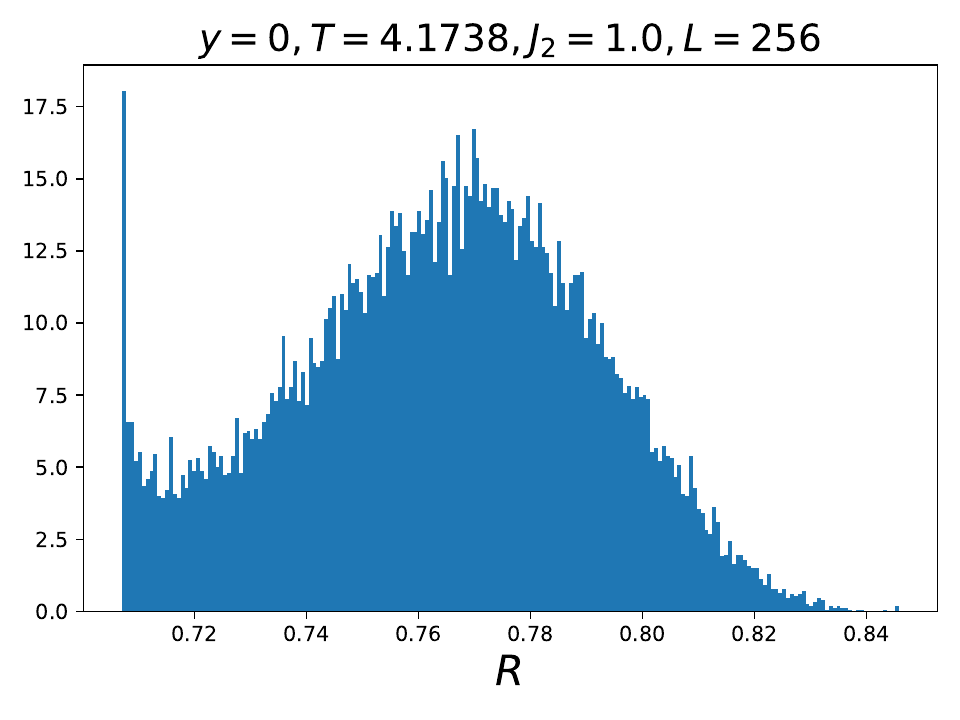}
			}
			\hbox{~~~~~~~~~	\includegraphics[width=0.3\textwidth]{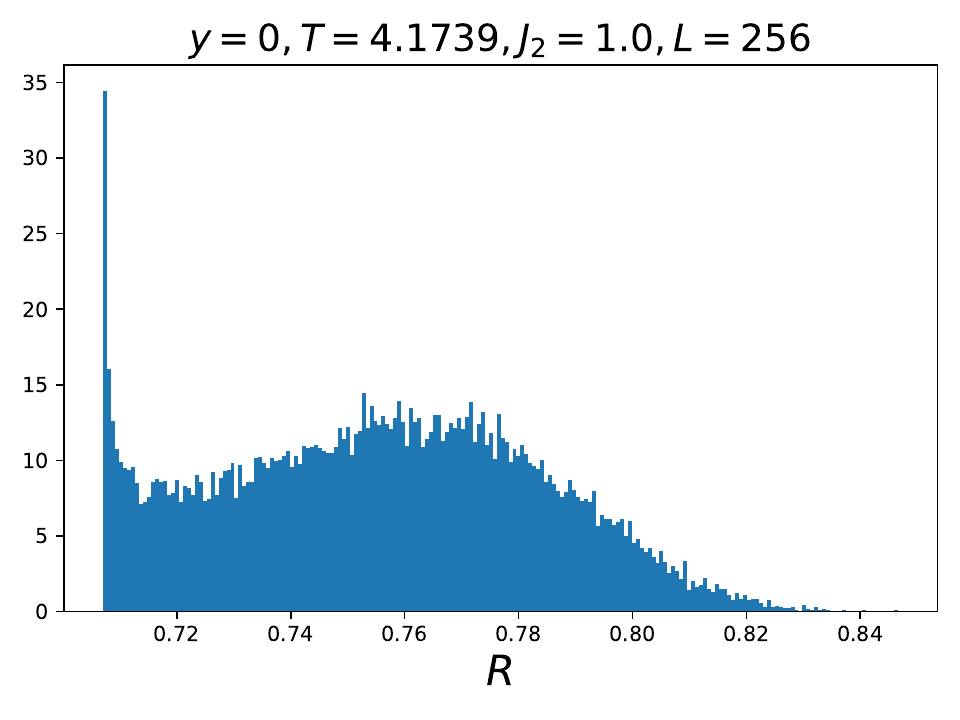}
				\includegraphics[width=0.3\textwidth]{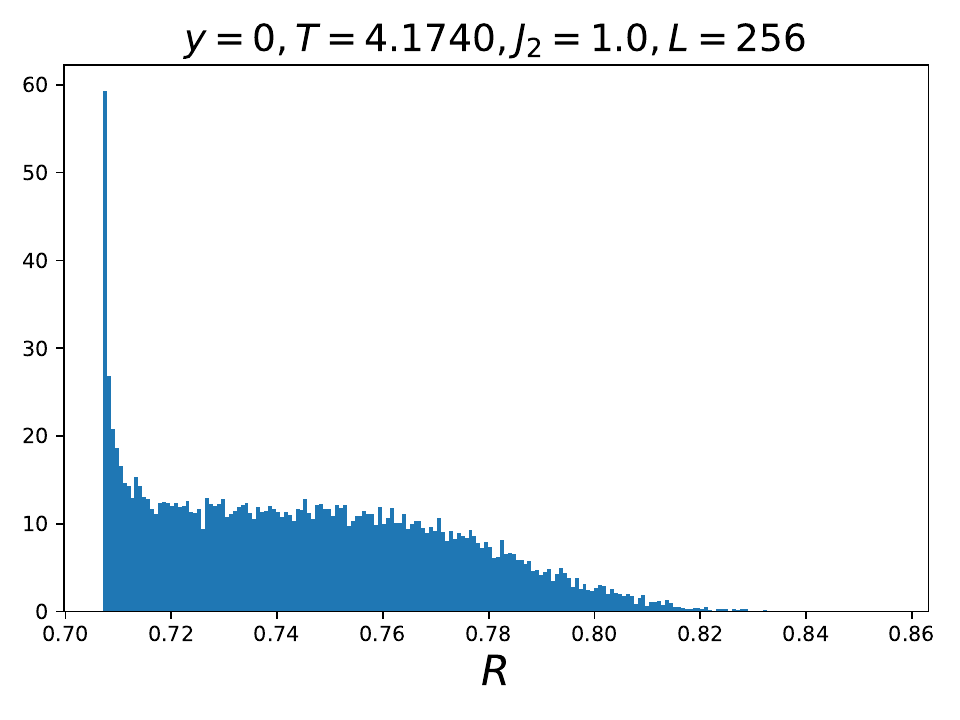}
				\includegraphics[width=0.3\textwidth]{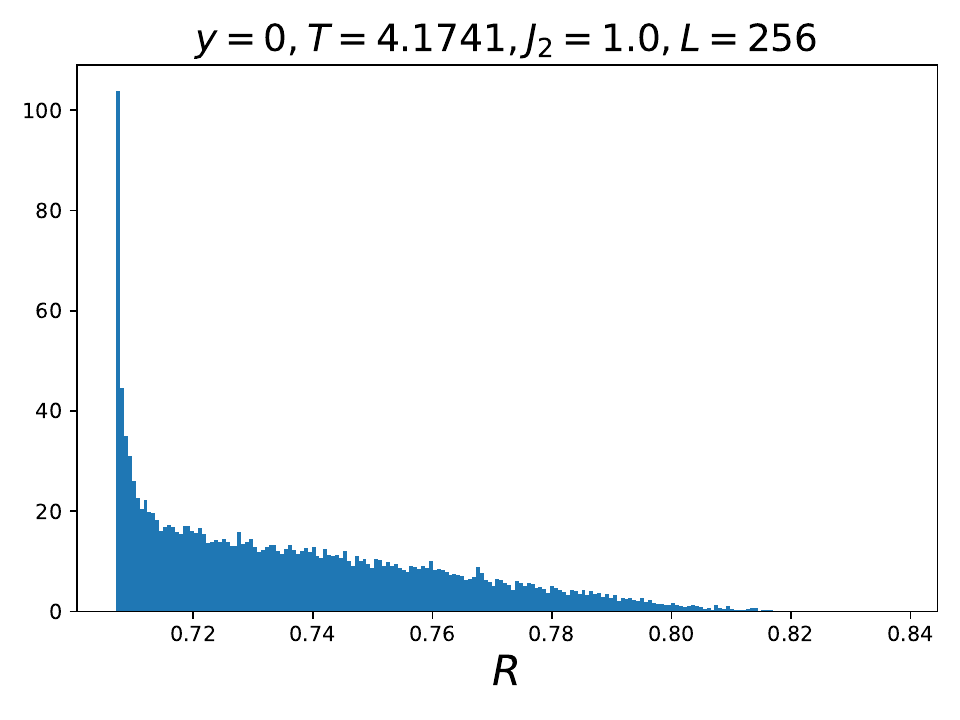}
			}
		}
	\end{center}
	\caption{Histograms of $R$ for $J_2 = 1.0$, $y=0$ and $L=256$ for several $T$ below and above $T_c$.
	}\label{RJ11zabovebelowL256}
\end{figure}

Figure \ref{RJ11zabovebelow} shows the histograms of $R$ for $J_2 = 1.0$ and $L=64$ at several $T$ close to the $T_c$ determined by MC data. The outcomes are obtained using the configurations with $z=0$. The behaviors of the histograms 
crossing $T_c$ are dramatically different from those associated with $J_2 = 4/7$. In particular,
no signals of coexisting phases are observed in Fig.~\ref{RJ11zabovebelow}. In other words, unlike the results shown in Fig.~\ref{potts5} which is associated with a weak first-order phase transition, the locations of the peaks are not close to 1 and the heights of 
the peaks are not dominated over that of the other values of $R$.
Moreover, a smooth evolution of histograms is found. These outcomes suggest that the phase transition for $J_2 = 1.0$ is unlikely first order.
Finally, the histograms of $R$ for $J_2 = 1.0$ and $L=256$ at several $T$ close to the $T_c$ computed by the MC method are depicted in Fig.~\ref{RJ11zabovebelowL256}, and the same messages as that of $L=64$ are obtained from Fig.~\ref{RJ11zabovebelowL256}. Based on these results, one arrives at the conclusion that the phase transition for $J_2 = 1.0$ is second order.

\begin{figure}
	
	\begin{center}
		\hbox{~~~~~~~~~~
			\includegraphics[width=0.4\textwidth]{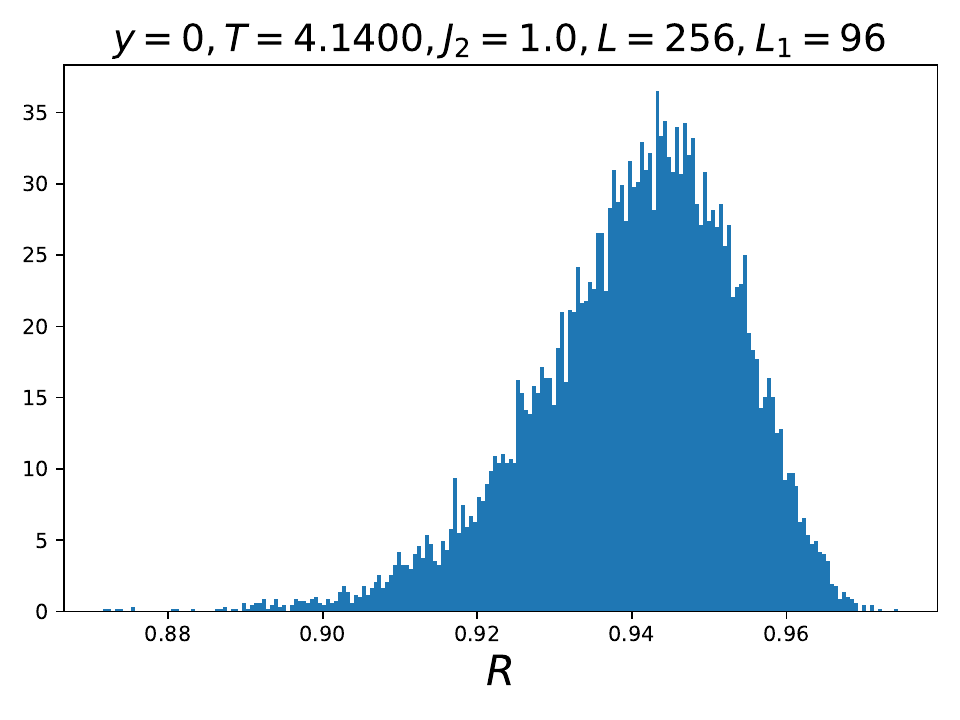}~~~~~~~
			\includegraphics[width=0.4\textwidth]{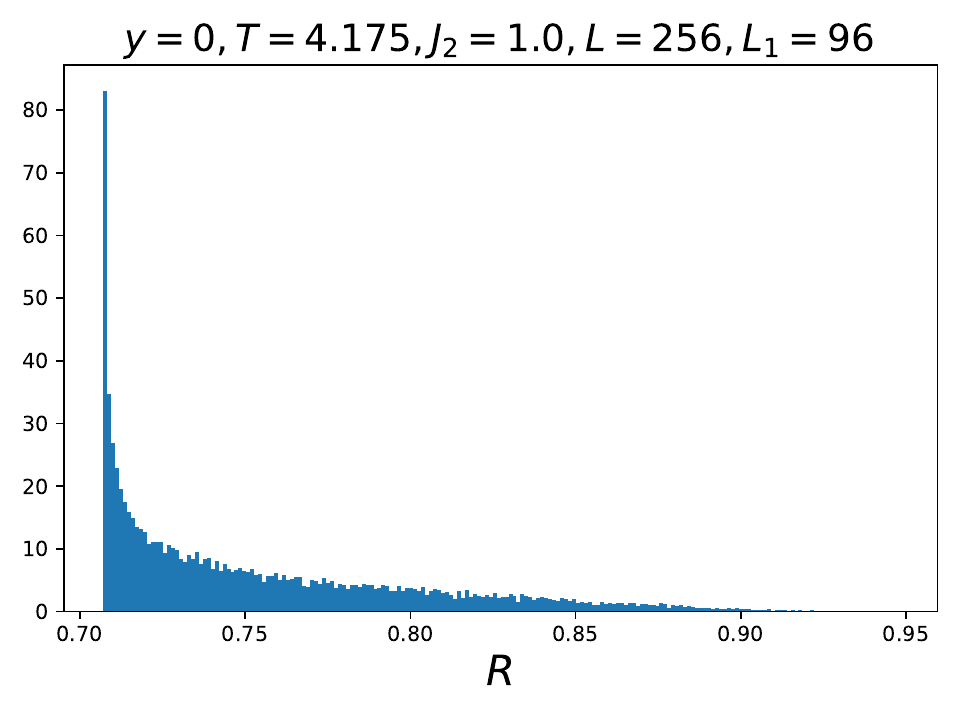}
		}
	\end{center}
	\caption{Histograms of $R$ for $J_2 = 1.0$, $y=0$, $L=256$ and $L_1=96$ for $T=4.1400$ (left) and $T = 4.1750$ (right).
	}\label{RJ11L96far}
\end{figure}

\begin{figure}

	\begin{center}
		\vbox{
			\hbox{~~~~~~~~
				\includegraphics[width=0.3\textwidth]{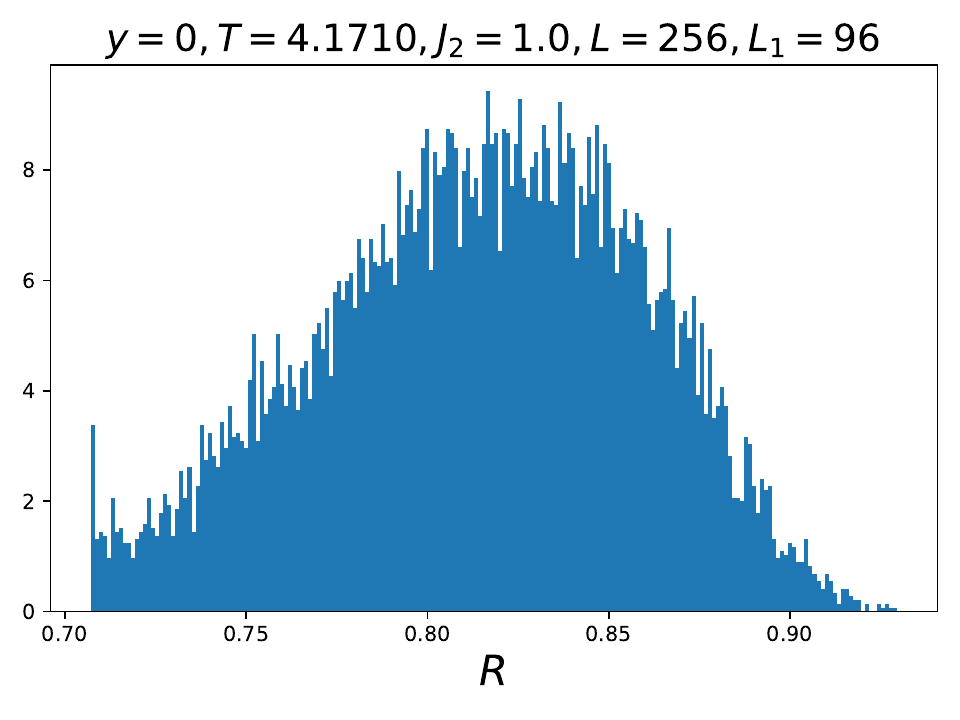}
				\includegraphics[width=0.3\textwidth]{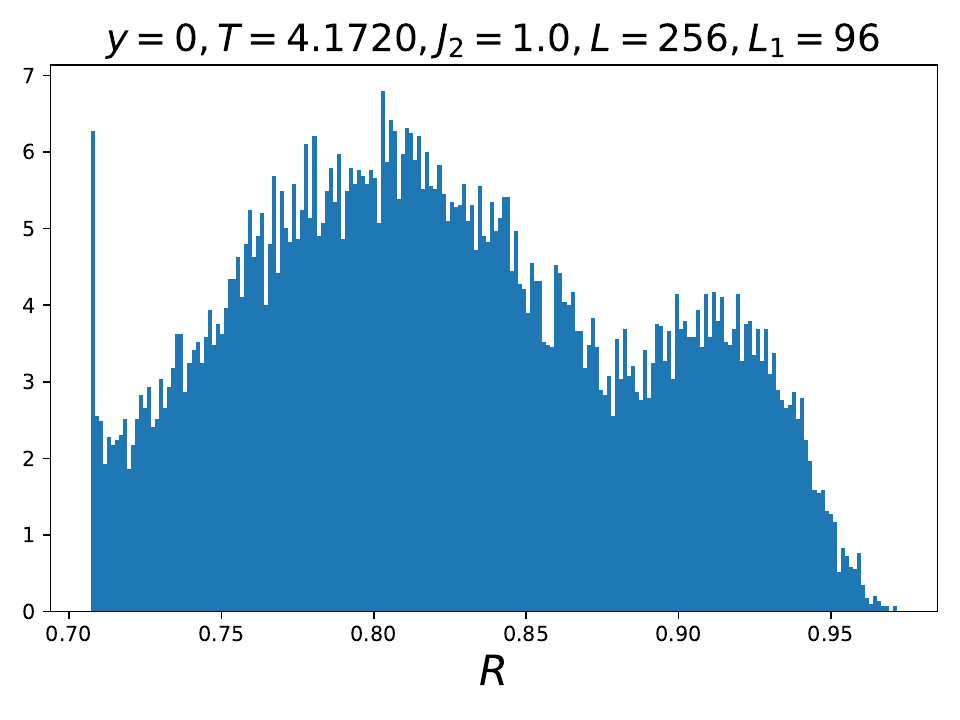}
				\includegraphics[width=0.3\textwidth]{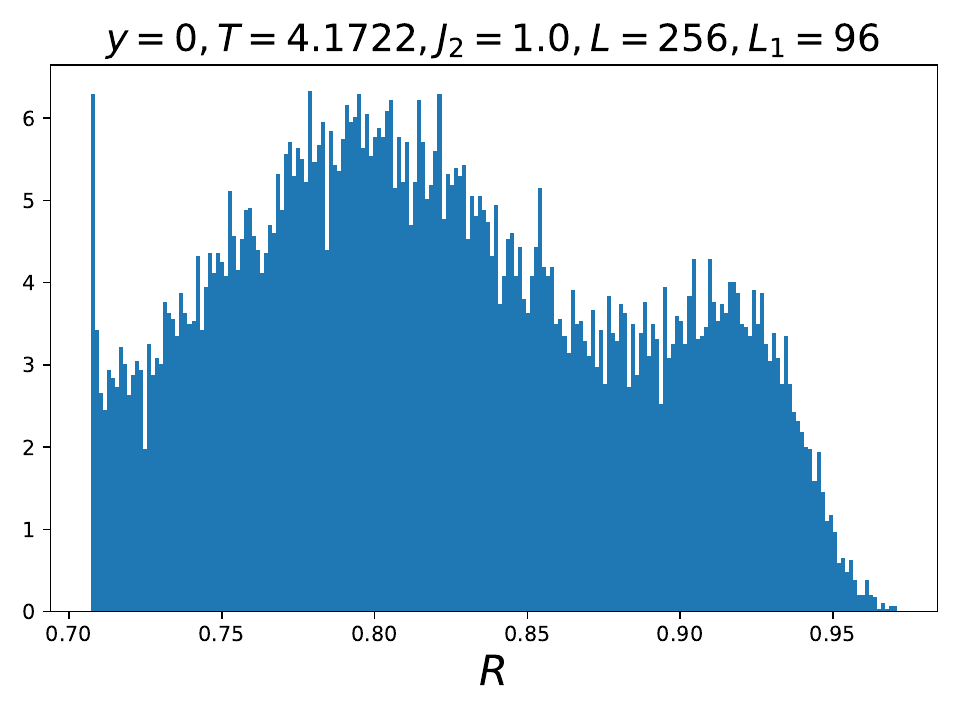}
			}
			\hbox{~~~~~~~~
				\includegraphics[width=0.3\textwidth]{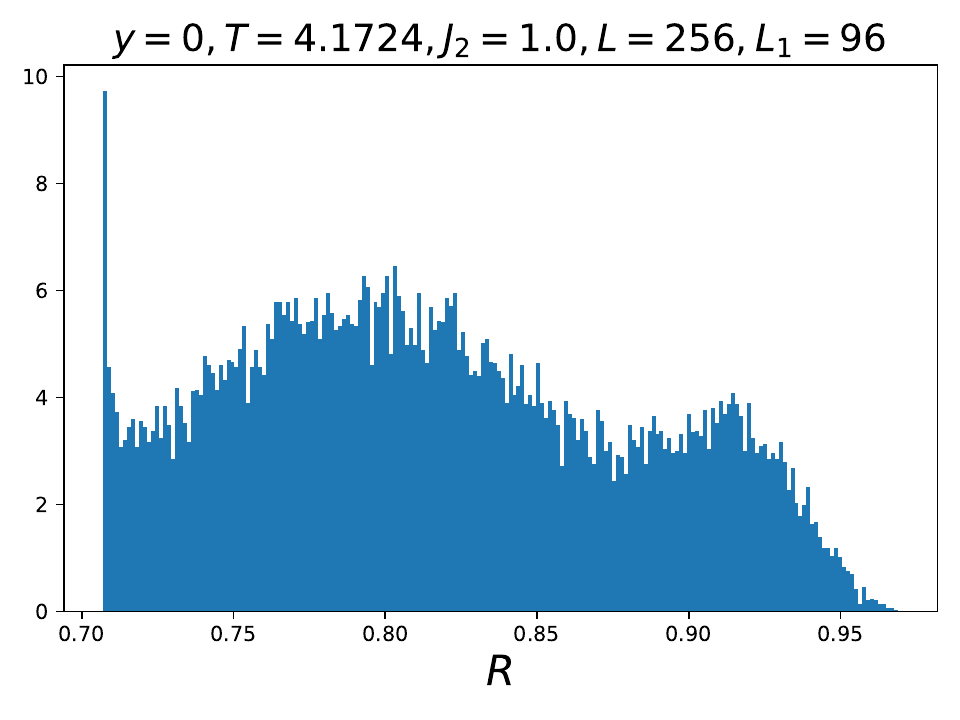}
				\includegraphics[width=0.3\textwidth]{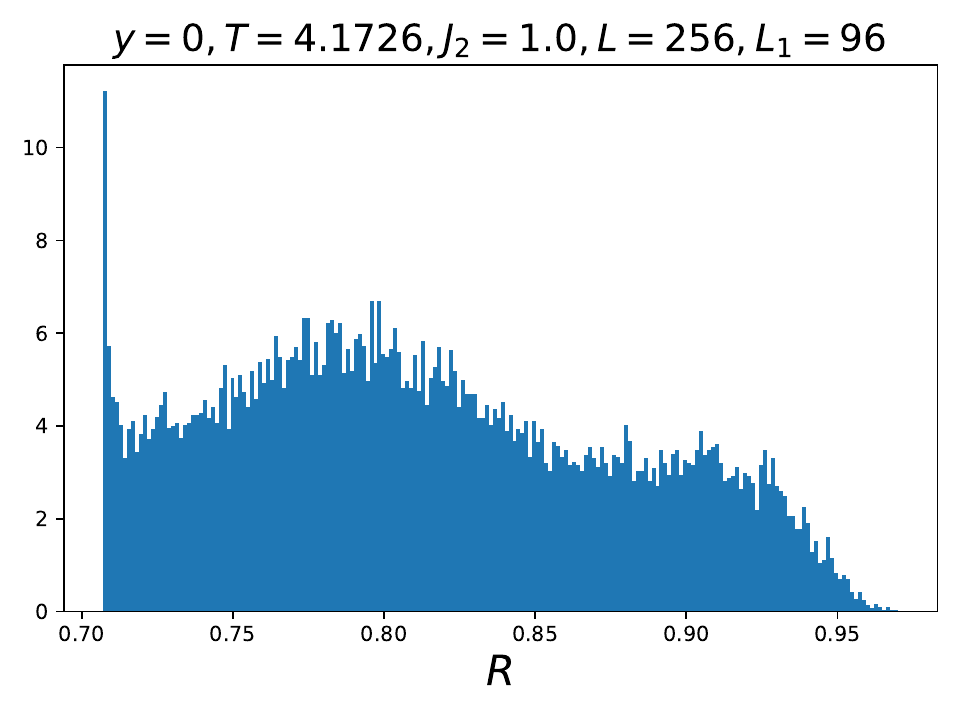}
				\includegraphics[width=0.3\textwidth]{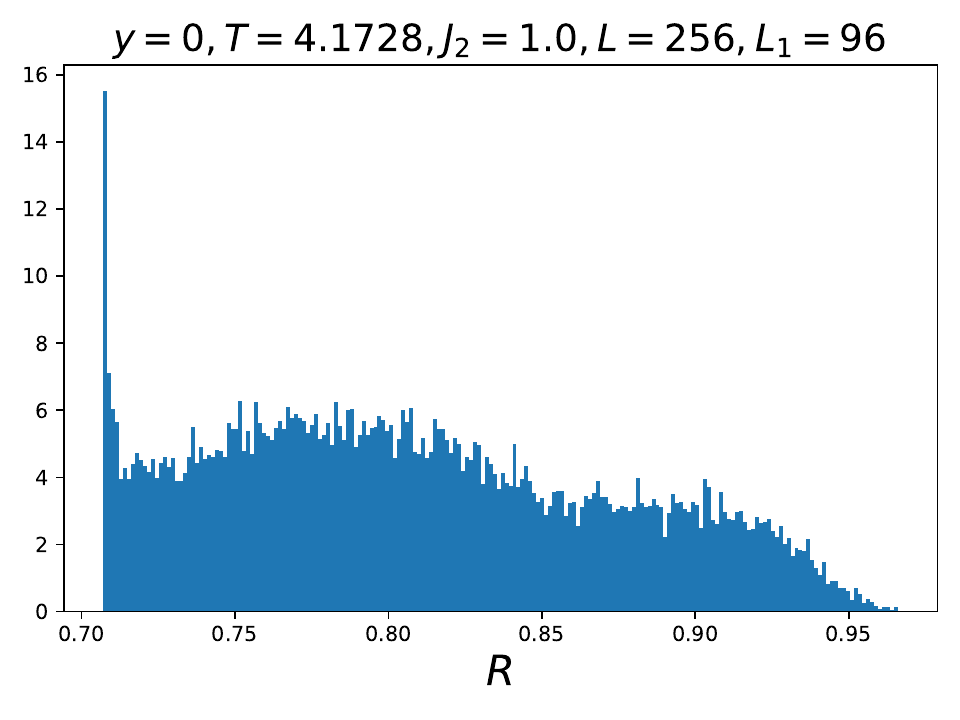}
			}
			\hbox{~~~~~~~~~	\includegraphics[width=0.3\textwidth]{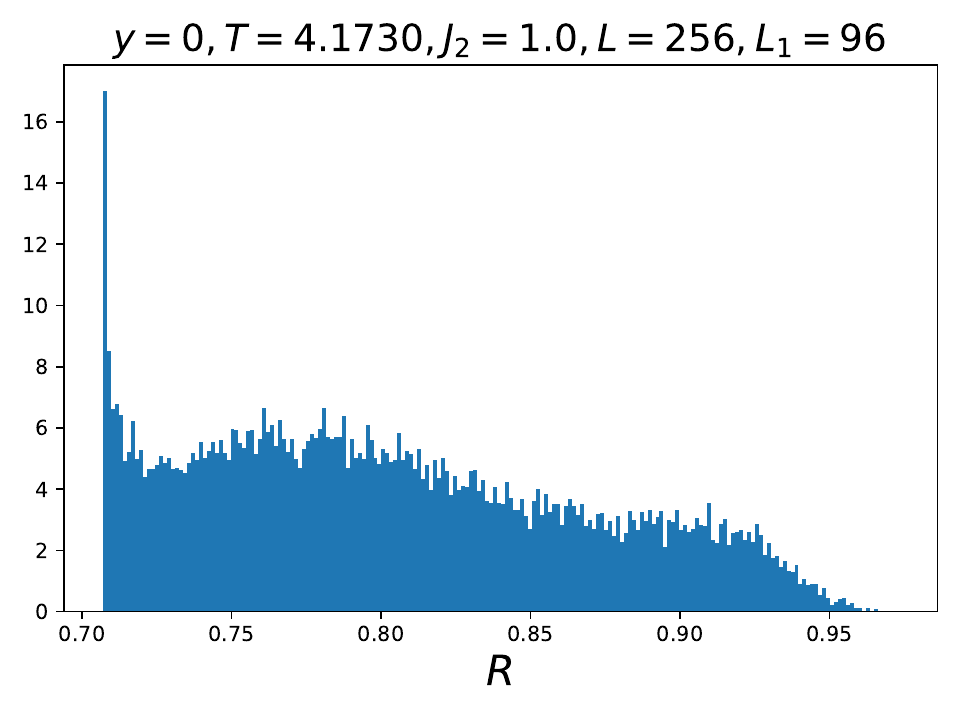}
				\includegraphics[width=0.3\textwidth]{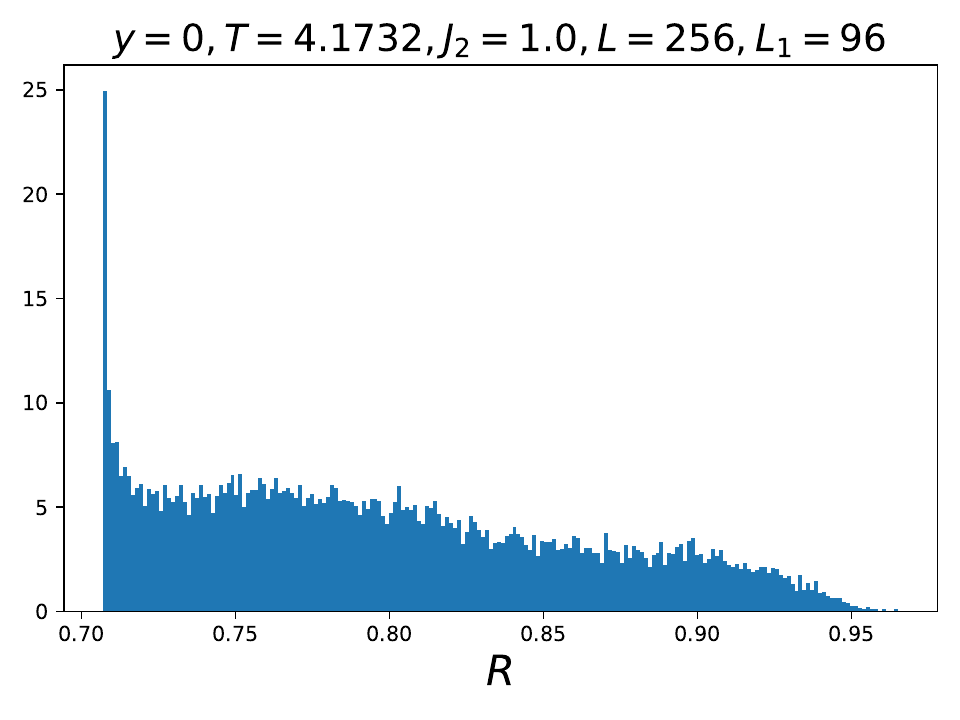}
				\includegraphics[width=0.3\textwidth]{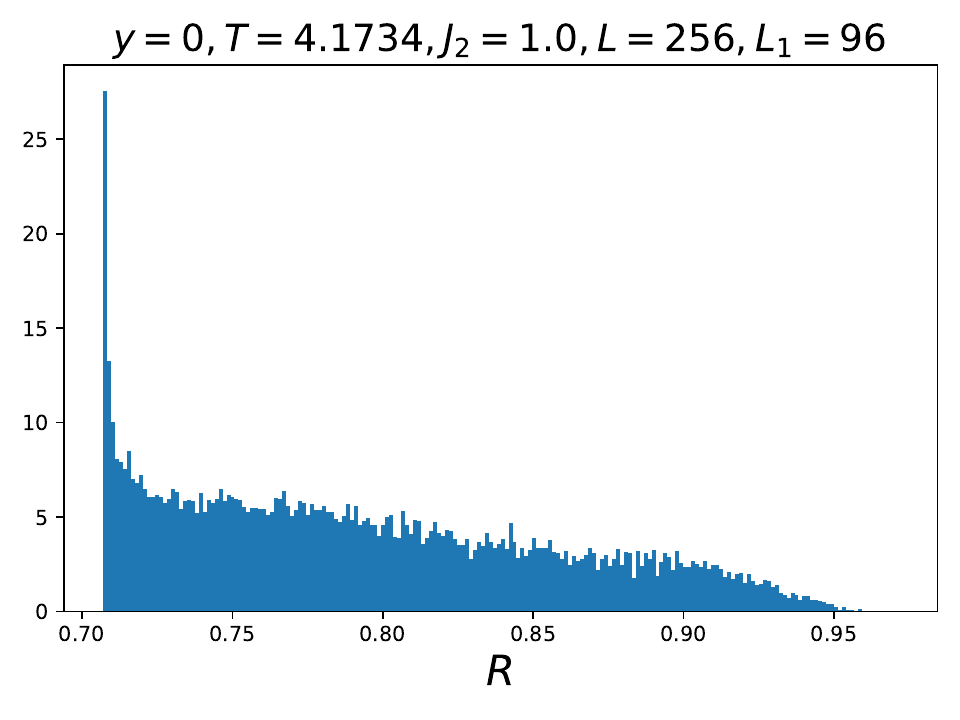}
			}
		}
	\end{center}
	\caption{Histograms of $R$ for $J_2 = 1.0$, $y=0$, $L=256$, and $L_1 = 96$ for several $T$ below and above $T_c$.
	}	\label{RJ1L256L96zabovebelow}
\end{figure}

Next, we would like to examine whether the use of the bottom left $L_1 \times L_1$ subsets of the whole recorded $256 \times 256$ spin 
configurations as the testing set can lead to a successful determination of the critical behaviors of the phase 
transition for $J_2 = 1.0$. Here we consider $L_1 = 96$. The left and right panels of Fig.~\ref{RJ11L96far} 
are the histograms of $R$ below and above the pseudo-critical temperature. The results presented
in Fig.~\ref{RJ11L96far} indicate that while the distributions of $R$ for $T$ below $T_c(L)$ show accumulation around a value, those above $T_c$ take values mainly near $1/\sqrt{2}$ and the distribution can spread out to a value $R\sim 0.9$. The panels of Fig.~\ref{RJ1L256L96zabovebelow} show
the histograms of $R$ for several $T$ below and above $T_c$. The results are obtained by only considering the bottom left $96 \times 96$ subsets of the whole recorded $256 \times 256$ configurations as the testing set. The outcomes given in Fig.~\ref{RJ1L256L96zabovebelow} imply the distributions of $R$ evolve continuously across $T_c(L)$. In particular, none of the panels of Fig.~\ref{RJ1L256L96zabovebelow} contain features of both $R$ histograms appearing in Fig.~\ref{RJ11L96far}. This suggests the histograms of $R$ obtained from the $96 \times 96$
subsets of the whole recorded $256 \times 256$ configurations favor the scenario that the phase transition of $J_2 = 1.0$ is likely second order.      

One may argue that a mild two-peak structure seems to show up in Fig.~\ref{RJ1L256L96zabovebelow}. Here we would like to emphasize the fact that for a fist-order phase transition, even a weak one, the associated histograms of $E$ close to $T_c(L)$ should contain both features of those corresponding to $T > T_c(L)$
and $T < T_c(L)$. For instance, the peaks in the histograms of $T=4.1710$ and $T=4.1734$ are located
at 0.825 and 0.7071, respectively. However, for all the three histograms of $T=4.1720,4.1722$ and $4.1724$, a peak of $R$ is found around $R \sim 0.925$. The scenario of the appearance of a peak around $R \sim 0.925$ does not take place neither at $T = 4.1710$ nor $T = 4.7134$. In summary, taking all the outcomes into considerations, the phase transition associated with $J_2 = 1.0$ is second order beyond reasonable doubt.

\subsection{Discussions and Conclusions}

Motivated by the controversial results regarding the nature of the phase transition related to $J_2 = 1.0$ 
shown in Refs.~\cite{Ram15,Sor18,Sor22}, in this study we perform an investigation of the 3D antiferromagnetic $J_1$-$J_2$-$J_3$ Ising model on the cubic lattice using both MC simulations and NN methods.
In particular, we have focused on the situations of $J_2 = 4/7$ and $J_2 = 1.0$ with $J_1 = J_3 = 1.0$.
Here $J_1$ and $J_2$ are the intralayer nearest-neighbor and next-nearest-neighbor couplings, respectively, and 
$J_3$ is the nearest-neighbor interlayer coupling.

In Refs.~\cite{Ram15,Sor18,Sor22}, the largest linear system size is $L \le 128$. By simulating an $L=256$ lattice, we find the phase transition of $J_2 = 1.0$ is second order based on the histograms of $E$. In other words, the controversy over the results presented in Refs.~\cite{Ram15,Sor18,Sor22} regarding the nature of the phase transition for $J_2 = 1.0$ (and $J_1 = J_3 =1.0$) is settled in this study. In addition, with the MC data, we confirm
that the phase transition of $J_2 = 4/7$ is first order. In particular, while it is well known
that the tunneling rate for a first-order phase transition in spatial dimension three is exponentially suppressed as $\exp(-c\beta L^{2})$, surprisingly we find that this scenario may not be applicable if the initial configurations
to start the MC simulations belong to the ground-state configurations. Examining whether this is indeed the case is worthwhile.

Here we do not perform calculations to compute the associated (unusual) critical exponents corresponding to $J_1 = J_2 = J_3 = 1.0$ since both Ref.~\cite{Ram15} and Ref.~\cite{Sor22} give the same numerical values for these exponents.

In our NN approach, the training procedure is conducted using four stripe and two N\'eel configurations on a square area with linear size $L$   
as the training set. Moreover, the testing sets consist of the subsets of the 3D spin configurations with 
either the coordinate $y=0$ or $z=0$. With such an approach, the required storage capacity and computational efficiency are greatly improved compared with those of conventional NN procedures. The validity of our
unconventional ideas is confirmed by the quantitative consistency of the values of $T_c$ and the nature of the targeted phase transitions obtained from the two different numerical methods
employed here.

In conclusion, with data from much larger lattices than those previously studied, we resolve the controversial results claimed in Refs.~\cite{Ram15,Sor18,Sor22}, namely, phase transition for $J_1 = J_2 = J_3 = 1.0$ 
is (likely) second order. Finally, the NN outcomes demonstrated here imply our NN procedure can provide a highly efficient approach to access very large lattices for systems of higher dimensions.

\section*{Acknowledgement}\vskip-0.3cm
Partial support from the National Science and Technology Council (NSTC) of 
Taiwan (NSTC 114-2112-M-003-004- and NSTC 115-2112-M-003-003-) is acknowledged.

\end{document}